\documentclass[pre,eqsecnum,tighten]{revtex4}

\usepackage{amssymb,amsmath}

\usepackage{epsfig}
\usepackage{color}

\usepackage{graphicx}

\usepackage{bm}

\begin{document}
\title{Enskog Theory for Polydisperse Granular Mixtures. III. Comparison
of dense and dilute transport coefficients and equations of state for a
binary mixture}
\author{J. Aaron Murray}
\author{Christine M. Hrenya}\email{hrenya@colorado.edu}
\affiliation{Department of Chemical and Biological Engineering,
University of Colorado, Boulder, CO 80309}
\author{Vicente Garz\'o}
\email{vicenteg@unex.es}
\homepage{http://www.unex.es/eweb/fisteor/vicente/} \affiliation{Departamento
de F\'{\i}sica, Universidad de Extremadura, E-06071 Badajoz, Spain}

\begin{abstract}
The objective of this study is to assess the impact of a dense-phase
treatment on the hydrodynamic description of granular, binary
mixtures relative to a previous dilute-phase treatment.  Two theories
were considered for this purpose. The first, proposed by Garz\'{o} and
Dufty (GD) [Phys. Fluids {\bf 14}, 146 (2002)], is based on the Boltzmann
equation which does not
incorporate finite-volume effects, thereby limiting its use to dilute
flows. The second, proposed by Garz\'{o}, Hrenya and Dufty (GHD) [Phys.
Rev. E {\bf 76}, 31303 and 031304 (2007)], is derived from the
Enskog equation which does account for finite-volume effects;
accordingly this theory can be applied to moderately dense systems as
well. To demonstrate the significance of the dense-phase
treatment relative to its dilute counterpart, the ratio of dense
(GHD) to dilute (GD) predictions of all relevant transport
coefficients and equations of state are plotted over a range of
physical parameters (volume fraction, coefficients of restitution,
material density ratio, diameter ratio, and mixture composition).
These plots reveal the deviation between the two treatments, which
can become quite large ($>$100\%) even at moderate values of the
physical parameters.  Such information will be useful when choosing
which theory is most applicable to a given situation, since the
dilute theory offers relative simplicity and the dense theory offers
improved accuracy.  It is also important to note that several
corrections to original GHD expressions are presented here in the
form of a complete, self-contained set of relevant equations.
\end{abstract}

\pacs{05.20.Dd, 45.70.Mg, 51.10.+y} \draft
\date{\today}
\maketitle

\section{Introduction}
\label{sec1}

Polydisperse, rapid solids flows are quite prevalent in both nature (i.e.,
landslides, avalanches)  and industry (i.e., pharmaceutical processing,
high-velocity fluidized beds), though much remains to be understood.
Perhaps most importantly, due to differences in size and/or material
density of each particle species, polydisperse mixtures are well-known to
exhibit particle segregation, also known as de-mixing \cite{S87,B94,OK00,RB02}.
Such behavior has no monodisperse counterpart. Thus, continuum
models developed for monodisperse flows cannot be used to predict the
segregation of unlike particles which occurs in polydisperse systems.
Consequently, an accurate continuum model of a polydisperse solids mixture
lends itself to a variety of non-trivial applications, such as the design
of coal gasifiers for energy production.

The scope of the current study pertains to binary mixtures of inelastic
grains (negligible fluid phase) engaging in instantaneous, binary
collisions (rapid flows). Numerous previous contributors have proposed
continuum theories for such systems (for recent review, see Ref.\ \cite{H10}),
and the application of these theories has led to a better understanding of
the mechanisms by which de-mixing occurs \cite{OK00,HH96,BRM05,GDH05,G06,
SUKSS06,WJKT06,YJ06,LMG07,G08,G09,GF09}.
Nevertheless, the improvement of existing models remains an active area of
research due to differences in the derivation process.  More specifically,
one or both of the following simplifications have been incorporated in the
vast majority of previous models: (i)~Maxwellian velocity distribution
\cite{JM87,MSAH99,HGM01,RNW03,IA05},
and/or (ii)~an equipartition of energy \cite{JM89,AW98}.
(The theories proposed by Rahaman {\it et al.} \cite{RNW03} as well as
Iddir and Arastoopour \cite{IA05} assumed a Maxwellian velocity distribution
between unlike particles only.)  The aforementioned assumptions
are strictly true for systems of perfectly elastic spheres in a uniform
steady state \cite{CC70}, but not so for inelastic grains.
Furthermore, numerous studies have demonstrated the influence of
non-equipartition on species segregation \cite{BRM05,GDH05,G06,YJ06,
LMG07,G08,G09}. Two continuum models have been proposed over the past decade
in which neither of the above conditions is assumed. The first theory,
developed by Garz\'{o} and Dufty (GD) \cite{GD02,GMD06,GM07} is based on the
Boltzmann equation,
and is thus applicable to dilute flows only.  The second theory, developed
recently by Garz\'{o}, Hrenya and Dufty (GHD) \cite{GDH07,GHD07}, instead
uses the Enskog equation as its starting
point, making this theory applicable to moderately dense systems as well.
Hereafter, the acronyms GD will be used to refer to the former, and GHD
will be used to refer to latter.  Each theory gives rise to a set of
zeroth-order closures known as equations of state, as well as constitutive
relations for first-order contributions to fluxes, or more specifically
the associated transport coefficients.
The equations of state and transport coefficients are functions of the
hydrodynamic variables:  number densities ($n_i$), mass-based mixture velocity
(${\bf U}$), and number-based mixture
granular temperature ($T$). Although the predictions
for the equations of state and transport coefficients from the two theories are
expected to match at low volume fractions, a non-negligible difference is
expected at higher concentrations, though the level of discrepancy between
the two has not yet been reported for polydisperse systems.

To build on the previous contributions, the focus of this work is to
analyze binary mixtures, where the two particle species differ in mass
and/or size. Motivation for this study is threefold: (i)~to assess
importance of dense-phase corrections to hydrodynamic description of
mixtures proposed by Garz\'{o} and co-workers \cite{GDH07,GHD07} compared
to the previous
dilute-phase description \cite{GD02}, and more specifically to determine
rules-of-thumb for the volume fraction at which such dense-phase
descriptions become non-negligible, (ii)~to examine the behavior of the
GHD equations of state and transport coefficients over a range of physical
parameters, and (iii)~to provide a complete, self-contained set of the GHD
expressions, including several corrections (see the Appendix \ref{appA}) for the expressions given in
the original GHD contribution \cite{GDH07,GHD07}. This latter goal also
provides an opportunity to display the expressions in a form more suitable
for computational purposes.

To accomplish the first two objectives, the equations of state and
transport coefficients were evaluated over a range of volume fractions,
coefficients of restitution, and mixture properties (diameter ratio, mass
ratio, and volume fraction ratio). The results indicate that the discrepancy
between transport coefficients and equations of state predicted by each
theory at a volume fraction of $\phi=0.1$ can vary from a factor of 1.05 to a
factor of 10. As the volume fraction becomes fairly dense ($\phi=0.5$), the
predicted discrepancy increases to a factor of at least 1.7 and as large
as a factor of 120. Hence, though the derivation of the constitutive
relations for a dilute flow and the resulting constitutive expressions are
simpler than its moderately dense counterpart, the difference between
the two theories is non-negligible at low to moderate volume fractions. In the
upcoming sections, a complete, self-contained set of the GHD constitutive
relations for the mass flux, heat flux, pressure tensor, and cooling rate
are given in Sections \ref{sec2}. Also, a quantitative comparison between
the GHD and GD predictions for the transport coefficients and equations of
state illustrates stark differences between the dilute and dense treatments
(Sections \ref{sec3} and \ref{sec4}). The paper is closed in Sec.\ \ref{sec5}
with a brief summary of the main results obtained here.

\section{Enskog kinetic theory for mass, momentum and heat fluxes and equations of state
of a granular binary mixture}
\label{sec2}

The mass, momentum, and granular energy balances for the GHD theory for an $s$-component mixture are given in Table 1,  along with the corresponding flux laws.
Each balance equation is expressed in terms of the hydrodynamic variables
($n_i$, ${\bf U}$, $T$), along with the following quantities:
cooling rate ($\zeta$), mass flux (${\bf j}_i$), heat flux (${\bf q}$),
and pressure  tensor (${\sf P}$). Constitutive expressions for these latter
quantities, also given in Table 1, are in terms of $\zeta^{(0)}$
(zeroth-order cooling rate), $\zeta_u$ (transport coefficient associated with
first-order cooling rate), $D_{ij}$ (mutual diffusion coefficients), $D_i^T$ (thermal
diffusion coefficients), $D_{ij}^F$ (mass mobility), $\lambda$ (thermal conductivity coefficient),
$D_{q,i}$ (Dufour coefficients), $L_{ij}$ (thermal mobility), $p$
(pressure), $\eta$ (shear viscosity), and $\kappa$ (bulk viscosity).
Also, ${\bf F}_i$ refers to the external force on a particle of species $i$.
To fully close the set of equations, these quantities must be cast in
terms of the hydrodynamic variables. The equations needed to obtain
closures for each expression are detailed in the following subsections;
corresponding equation numbers are also listed in Table 1.

\subsection{Mass and heat fluxes}

We consider a binary mixture ($s=2$) of {\em inelastic}, smooth, hard disks
($d=2$) or spheres ($d=3$) of masses $m_1$ and $m_2$, and diameters $\sigma_1$
and $\sigma_2$. The inelasticity of collision among all pairs is
characterized by three independent constant
coefficients of normal restitution $\alpha_{11}$, $\alpha_{22}$,
and $\alpha_{12}=\alpha_{21}$, where $\alpha_{ij}$ is the coefficient of
restitution for collisions between particles of species $i$ and $j$.
For moderate densities, it is assumed  that the
velocity distribution functions of each species are accurately
described by the coupled set of {\em inelastic} Enskog kinetic equations
\cite{GS95,BDS97}.
This set of equations has been recently solved in Refs.\ \cite{GDH07,
GHD07} by means of the Chapman-Enskog method \cite{CC70} and the constitutive
equations for
the mass ${\bf j}_1$ and heat ${\bf q}$ fluxes have been obtained up to
the Navier-Stokes order (first-order in the spatial gradients). In the
absence of external forces (${\bf F}_i={\bf 0}$), the forms of ${\bf j}_1$ and ${\bf q}$ are given,
respectively, by
\begin{equation}
\mathbf{j}_{1}=-\frac{m_1^2n_{1}}{\rho} D_{11}\nabla \ln n_{1}-
\frac{m_1 m_2n_{2}}{\rho} D_{12}\nabla \ln n_{2}
-\rho D_{1}^{T}\nabla \ln T,  \label{2.1}
\end{equation}\vspace*{-.1in}
\begin{equation}
\label{2.2}
{\bf q}=-T^2D_{q,1}\nabla \ln n_1-T^2 D_{q,2}\nabla \ln n_2-\lambda
\nabla T,
\end{equation}
where $\rho=m_1n_1+m_2n_2$ and $n_i$ refers to the number density of
species $i$. While the diffusion coefficients $D_{ij}$ and the thermal
diffusion coefficient $D_i^T$ have only kinetic contributions, the
transport coefficients $D_{q,i}$ and $\lambda$ associated with the
heat flux have also collisional transfer contributions. Expressions
for these transport coefficients in terms of
the coefficients of restitution, the parameters of the mixture (masses,
sizes and composition), and concentration (solid volume fraction) have
been obtained in Ref.\ \cite{GHD07} by using the leading terms in a
Sonine polynomial expansion.

\begin{center}
\label{tab1}
{\bf Table 1:} Hydrodynamic description of a granular mixture from GHD theory.
\bigskip

\def\baselinestretch{2}\large\normalsize
\begin{tabular}{|l|l|}\hline\hline
\multicolumn{2}{|l|} {\bf Balance Equations}\\ \hline\hline
Mass & $\displaystyle\frac{Dn_i}{Dt} + n_i\nabla\cdot
{\bf U}=-\displaystyle\frac{1}{m_i}\nabla\cdot
{\bf j}_{i}$\\ \hline
Momentum & $\rho\displaystyle\frac{D{\bf U}}{Dt} =-\nabla\cdot{\bf P}+\sum_{i=1}^s\;
n_i{\bf F}_i$
\\ \hline
Granular &
$\displaystyle\frac{d}{2}n\displaystyle\frac{DT}{Dt}=-\nabla\cdot
{\bf q}-{\bf P}:\nabla {\bf U}-\displaystyle\frac{d}{2}nT\zeta+
\displaystyle\frac{d}{2}T
\displaystyle\sum_{i=1}^s\displaystyle\frac{1}{m_i}\nabla\cdot
{\bf j}_{i}+\sum_{i=1}^s\;\frac{{\bf F}_i\cdot {\bf j}_i}{m_i}$\hspace*{.15in}\vspace*{-.2in}\\
Energy & \\ \hline\hline \multicolumn{2}{|l|}{}\vspace*{-.2in}\\ \hline\hline
\multicolumn{2}{|l|}{\bf Flux Laws}\\ \hline
Mass & ${\bf j}_i=-\displaystyle\sum_{j=1}^s\;m_im_j
\displaystyle\frac{n_j}{\rho}D_{ij}\nabla\ln n_j-\rho D_i^T
\nabla\ln T-\sum_{j=1}^s\;D_{ij}^F{\bf F}_j$
\hfill Eq.\ 2.1\\ \hline
Heat & ${\bf q}=-\lambda \nabla T-\displaystyle\sum_{i=1}^sT^2D_{q,i}
\nabla\ln n_i-\sum_{i=1}^s\sum_{j=1}^s\;L_{ij}{\bf F}_i$\hfill Eq.\ 2.2\\ \hline
Pressure & $P_{\alpha\beta}=p\delta_{\alpha\beta}-\eta\left(\nabla_\alpha U_\beta+
\nabla_\beta U_\alpha
-\frac{2}{d} \nabla\cdot{\bf U}\delta_{\alpha\beta}\right)
-\kappa \nabla\cdot{\bf U}\delta_{\alpha\beta}$
\hfill Eq.\ 2.22\\ \hline\hline
\multicolumn{2}{|l|}{}\vspace*{-.2in}\\ \hline\hline
{\bf Cooling Rate}\ \ \ \ & $\zeta=\zeta^{(0)}+\zeta_u\bf{\nabla}\cdot{\bf U}$
\\ \end{tabular}

\begin{tabular}{|l|l|l||l|l|l|}\hline\hline
\multicolumn{3}{|c||}{\bf Equations of State}
& \multicolumn{3}{|c|} {\bf Transport Coefficient}\\ \hline\hline
Zeroth-Order & $\zeta^{(0)}$ \ \ &
Eq.\ 2.38  & First-Order  & $\zeta_u$ & \hfill
Eqs.\ 2.40, 2.41, 2.42\vspace*{-.15in}\\
Cooling Rate & & & Cooling Rate & & \\ \hline
Pressure & $p$ & Eqs.\ 2.24, 2.25 & Shear Viscosity & $\eta$ & \hfill
Eqs.\ 2.27, 2.28, 2.30, 2.35\\ \hline
\multicolumn{3}{|c||} {} & Bulk Viscosity & $\kappa$ & \hfill Eqs.\ 2.34, 2.36\\
\cline{4-6}
\multicolumn{3}{|c||}{} & Mutual Diffusion & $D_{ij}$ & \hfill Eqs.\ 2.3, 2.7, 2.8\\ \cline{4-6}
\multicolumn{3}{|c||}{} & Thermal Diffusion  & $D_i^T$ & \hfill Eqs.\
2.3, 2.6\\ \cline{4-6}
\multicolumn{3}{|c||}{} & Thermal conductivity & $\lambda$ & \hfill Eqs.\ 2.4, 2.13, 2.16, 2.17\\ \cline{4-6}
\multicolumn{3}{|c||}{} & Dufour & $D_{q,i}$ & \hfill \ \ \ \ Eqs.\ 2.4, 2.13, 2.15, 2.20 \\ \cline{4-6}
\multicolumn{3}{|c||}{} & Mass Mobility & $ D_{ij}^F$\ \  & \\ \cline{4-5}
\multicolumn{3}{|c||}{} & Thermal Mobility & $L_{ij}$ & \\ \hline
\end{tabular}\end{center}\bigskip

The transport coefficients depend upon the temperature,
concentration, and composition of the mixture, as well as the masses,
diameters, and coefficients of restitution. To present the expressions
of these coefficients in a compact form, it is convenient to consider
their dimensionless forms:
\begin{equation}
\label{2.3}
D_{11}^*=\frac{m_1^2\nu_0}{\rho T}D_{11}, \quad D_{12}^*=\frac{m_1m_2\nu_0}
{\rho T}D_{12}, \quad
D_1^{T*}=\frac{\rho \nu_0}{nT}D_1^T,
\end{equation}
\begin{equation}
\label{2.4}
D_{q,i}^*=\frac{2}{d+2}\frac{(m_1+m_2)\nu_0}{n}D_{q,i}, \quad \lambda^*=
\frac{2}{d+2}\frac{(m_1+m_2)\nu_0}{n T}
\lambda.
\end{equation}
Here, $\nu_0=n\sigma_{12}^{d-1}v_0$ is an effective collision frequency,
$\sigma_{12}=(\sigma_1+\sigma_2)/2$,
and $v_0=\sqrt{2T/m}$ is a thermal velocity where $m=(m_1+m_2)/2$.
Thus, the results given throughout the remainder of the paper are given
in terms of the mass ratio $m_1/m_2$,
the size ratio $\sigma_1/\sigma_2$, the species number fraction
$x_i=n_i/n$, the
concentration $n\sigma_2^{d}$, and the coefficients of restitution
$\alpha_{11}$, $\alpha_{22}$, and $\alpha_{12}$.

\subsection*{A1. Mass flux transport coefficients}
\label{subA1}

In a binary mixture, since $\mathbf{j}_{1}=-\mathbf{j}_2$,
the mass flux contains three relevant transport coefficients: $D_{11}$,
$D_{12}$, and $D_{1}^T$. The remaining coefficients $D_{22}$, $D_{21}$,
and $D_{2}^T$ are given by the identities
\begin{equation}
\label{2.5}
D_{21}=-\frac{m_1}{m_2}D_{11}, \quad D_{22}=-\frac{m_1}{m_2}D_{12},
\quad D_2^T=-D_1^T.
\end{equation}
The expressions of the reduced coefficients $D_{1}^{T*}$, $D_{11}^*$,
and $D_{12}^*$ can be written as \cite{GHD07}
\begin{eqnarray}
D_{1}^{T*}&=&\left(\nu_D^*-\zeta^*\right)^{-1}\Big\{x_1\gamma_1-\frac{p^*
\rho_1}{\rho}+\frac{\pi ^{d/2}}{2d\Gamma \left( \frac{d}{2}\right)}x_1n \sigma_2^{d}
\left[x_1\chi_{11}(\sigma_1/\sigma_2)^{d}\gamma_1(1+\alpha_{11})
\right.\nonumber\\
& & \left.+2x_2\chi_{12}(\sigma_{12}/\sigma_2)^{d}M_{12}\gamma_2
(1+\alpha_{12})\right]\Big\},  \label{2.6}
\end{eqnarray}
\begin{eqnarray}
\label{2.7}
\left(\nu_D^*-\frac{1}{2}\zeta^*\right)D_{11}^*&=&\frac{D_{1}^{T*}}
{x_1\nu_0}n_1\frac{\partial \zeta^{(0)}}{\partial n_1}-\frac{m_1}{\rho T}
n_1\frac{\partial p}{\partial n_1}+\gamma_1+n_1\frac{\partial \gamma_1}{\partial n_1}
\nonumber\\
& & +\frac{\pi^{d/2}}{d\Gamma \left( \frac{d}{2}\right)}
x_1 n\sigma_2^{d}\sum_{\ell=1}^{2}\chi_{1\ell}(\sigma _{1\ell}/\sigma_2)^{d}
M_{\ell 1}(1+\alpha_{1\ell })\nonumber\\
& & \times\left\{ \frac{1}{2}\left(\gamma_1+\frac{m_1}{m_{\ell}}\gamma_\ell\right)
\left[ 2 \delta _{1\ell }+n_\ell
\frac{\partial \ln \chi_{1\ell}}{\partial n_1}+\frac{n_\ell}{n_1}I_{1\ell 1}
\right]+\frac{m_1}{m_\ell}n_\ell
\frac{\partial \gamma_\ell}{\partial n_1}\right\},\nonumber\\
\end{eqnarray}
\begin{eqnarray}
\label{2.8}
\left(\nu_D^*-\frac{1}{2}\zeta^*\right)D_{12}^*&=&\frac{D_{1}^{T*}}
{x_2\nu_0}n_2\frac{\partial \zeta^{(0)}}{\partial n_2}-\frac{m_1}
{\rho T}n_1\frac{\partial p}{\partial n_2}+n_1
\frac{\partial \gamma_1}{\partial n_2}\nonumber\\
& & +\frac{\pi^{d/2}}{d\Gamma \left( \frac{d}{2}\right)}
x_1 n\sigma_2^{d}\sum_{\ell=1}^{2}\chi_{1\ell}(\sigma _{1\ell}/\sigma_2)^{d}
M_{\ell 1}(1+\alpha_{1\ell })\nonumber\\
& & \times\left\{ \frac{1}{2}\left(\gamma_1+\frac{m_1}{m_{\ell}}\gamma_\ell\right)\left[ 2
\delta _{2\ell }+n_\ell
\frac{\partial \ln \chi_{1\ell}}{\partial n_2}+\frac{n_\ell}{n_2}I_{1\ell 2}\right]+\frac{m_1}
{m_\ell}n_\ell\frac{\partial \gamma_\ell}{\partial n_2}\right\}.\nonumber\\
\end{eqnarray}
In these equations, $\rho_i=m_in_i$, $\gamma_i=T_i/T$, $\zeta^*=\zeta^{(0)}/\nu_0$, $p^*=p/
(nT)$, $\chi_{ij}$ is the pair distribution function at contact, $M_{ij}=m_i/
(m_i+m_j)$, and
\begin{equation}
\label{2.9}
\nu_D^*= \frac{\sqrt{2}\pi^{(d-1)/2}}{d\Gamma \left( \frac{d}{2}\right)}\chi_{12}(1+\alpha_{12})\left(\frac{\gamma_1}
{M_{12}}+\frac{\gamma_2}{M_{21}}\right)^{1/2}(x_1 M_{12}+x_2 M_{21}).
\end{equation}
The partial temperatures $T_1$ and $T_2$ are determined from the condition
$\zeta_1^{(0)}=\zeta_2^{(0)}=\zeta^{(0)}$, where the expression of $\zeta_i^{(0)}$ is given by
Eq.\ (\ref{4.9}). Moreover, an explicit form for $\chi_{ij}$ for disks ($d=2$)
and spheres ($d=3$) is given in the Appendix \ref{appB}.

The parameters $I_{i\ell j}$ are chosen to recover the results derived by
L\'opez de Haro {\em et al.} for elastic mixtures \cite{MCK83}. These
quantities are the origin of the primary difference between the
standard Enskog theory and the revised version for elastic collisions \cite{RET}. They are zero if
$i=\ell$, but otherwise are not zero. They are defined through the relation \cite{GHD07}
\begin{equation}
\label{2.10} \sum_{\ell=1}^2 n_\ell\sigma_{i\ell}^d \chi_{i\ell}\left(n_j
\frac{\partial\ln \chi_{i\ell}}{\partial n_j}+I_{i\ell j}\right)=
\frac{n_j}{T B_2}\left(\frac{\partial \mu_i}{\partial n_j}\right)_{T,n_{k\neq
j}}-\frac{\delta_{ij}}{B_2}-2 n_j\chi_{ij}\sigma_{ij}^d.
\end{equation}
where $B_2=\pi ^{d/2}/d\Gamma \left(d/2\right)$
[$\Gamma$ refers to Gamma function, such that $B_2=\frac{\pi}{2}$
for $d=2$ (disks) and $B_2=2\pi/3$ for $d=3$ (spheres)]
and $\mu_i$ is the chemical potential of species $i$. Taking into account
Eq.\ (\ref{2.10}), the nonzero parameters $I_{121}$ and $I_{122}$ appearing in Eqs.\ (2.11) and (2.12) are given by
\begin{equation}
\label{a8}
I_{121}=\frac{1}{TB_2n_2\sigma_{12}^d\chi_{12}}\left[n_1\left(\frac{\partial \mu_1}{\partial n_1}
\right)_{T,n_{2}}-T\right]-2\frac{n_1\sigma_1^d\chi_{11}}
{n_2\sigma_{12}^d\chi_{12}}-
\frac{n_1^2\sigma_1^d}{n_2\sigma_{12}^d\chi_{12}}\frac{\partial\chi_{11}}{\partial n_1}
-\frac{n_1}{\chi_{12}}\frac{\partial\chi_{12}}{\partial n_1},
\end{equation}
\begin{equation}
\label{2.12}
I_{122}=\frac{1}{TB_2\sigma_{12}^d\chi_{12}}\left(\frac{\partial \mu_1}{\partial n_2}
\right)_{T,n_{1}}-2-
\frac{\sigma_1^d n_1}{\sigma_{12}^d\chi_{12}}\frac{\partial\chi_{11}}{\partial n_2}
-\frac{n_2}{\chi_{12}}\frac{\partial\chi_{12}}{\partial n_2}.
\end{equation}
Explicit forms of $\mu_i$ for disks ($d=2$) and spheres ($d=3$) are given in the Appendix \ref{appB}.

\subsection*{A2. Heat flux transport coefficients}
\label{subA2}

The heat flux requires going up to the second Sonine approximation. Its
constitutive equation is given by Eq.\ (\ref{2.2}) where the transport coefficients $D_{q,i}$ and $\lambda$ have kinetic and collisional contributions
\begin{equation}
\label{2.13}
D_{q,i}=D_{q,i}^k+D_{q,i}^c, \quad \lambda=\lambda^k+\lambda^c.
\end{equation}
The corresponding reduced forms $D_{q,i}^{k*}$, $D_{q,i}^{c*}$, $\lambda^{k*}$, and $\lambda^{c*}$ are defined as
\begin{equation}
\label{2.14}
D_{q,i}^{k,c*}=\frac{2}{d+2}\frac{(m_1+m_2)\nu_0}{n}D_{q,i}^{k,c}, \quad \lambda^{k,c*}=\frac{2}{d+2}\frac{(m_1+m_2)\nu_0}{n T}
\lambda^{k,c}.
\end{equation}

The kinetic parts $D_{q,i}^{k*}$ and $\lambda^{k*}$ can be written, respectively, as
\begin{equation}
\label{2.15}
D_{q,1}^{k*}=d_{q,11}^*+d_{q,21}^*+\left(\frac{\gamma_1}{M_{12}}-\frac{\gamma_2}{M_{21}}\right)
x_1 D_{11}^*, \quad
D_{q,2}^{k*}=d_{q,22}^*+d_{q,12}^*+\left(\frac{\gamma_1}{M_{12}}-\frac{\gamma_2}{M_{21}}\right)
x_2 D_{12}^*,
\end{equation}
\begin{equation}
\label{2.16}
\lambda^{k*}=\lambda_1^*+\lambda_2^*+\left(\frac{\gamma_1}{M_{12}}-\frac{\gamma_2}{M_{21}}\right)D_1^{T*},
\end{equation}
where the expressions for the (dimensionless) coefficients $d_{q,ij}^*$
and $\lambda_i^*$ are displayed in the Appendix \ref{appB}. In Eqs.\ (\ref{2.15})
and (\ref{2.16}), the coefficients $D_1^{T*}$, $D_{11}^*$ and $D_{12}^*$ are given by Eqs.\ (\ref{2.6}), (\ref{2.7}), and (\ref{2.8}), respectively (first Sonine approximation).

Let us consider now their collisional transfer contributions. In the case of the thermal conductivity, $\lambda^{c*}$ is given by \cite{GHD07}
\begin{eqnarray}
\label{2.17}
\lambda^{c*}&=&\frac{3}{2}\frac{\pi^{d/2}}{d(d+2)\Gamma \left(\frac{d}{2}\right)}n\sigma_2^d
\sum_{i=1}^2\sum_{j=1}^2
x_i(\sigma_{ij}/\sigma_2)^d\chi_{ij}M_{ij}(1+\alpha_{ij})\left\{\left[(5-\alpha_{ij})M_{ij}-(1-\alpha_{ij})
M_{ji}\right]\lambda_{j}^*\right.
\nonumber\\
& &  +(m_1+m_2)D_j^{T*}\left[\frac{\gamma_j}{m_j}\left((5-\alpha_{ij})M_{ij}-(1-\alpha_{ij})M_{ji}\right)
+\frac{\gamma_i}{m_i}\left((3+\alpha_{ij})M_{ji}-(7+\alpha_{ij})M_{ij}\right)\right]\nonumber\\
& &
\left.+\frac{16}{3\sqrt{\pi}}\frac{x_jm_j}{m_1+m_2}(\sigma_{12}/\sigma_2)^{d}(\sigma_{ij}/\sigma_{12})
C_{ij}^*\right\},
\end{eqnarray}
where
\begin{eqnarray}
C_{ij}^{*} &=&(\theta _{i}+\theta
_{j})^{-1/2}(\theta _{i}\theta
_{j})^{-3/2} \left\{ 2\beta _{ij}^{2}+\theta _{i}\theta _{j}+(\theta _{i}+\theta
_{j})\left[ (\theta _{i}+\theta _{j})M_{ij}M_{ji}+\beta_{ij}(1+M
_{ji})\right] \right\}  \notag \\
&&+\frac{3}{4}(1-\alpha _{ij})(M_{ji}-M
_{ij})\left( \frac{\theta_{i}+\theta_{j}}{\theta_{i}\theta_{j}}\right)^{3/2}\left[ M
_{ji}+\beta _{ij}(\theta _{i}+\theta _{j})^{-1}\right].  \notag \\
&&  \label{2.19}
\end{eqnarray}
Here, $\theta_i=(m_iT/mT_i)$ and $\beta_{ij}=M_{ij}\theta_j-M_{ji}\theta_i$.
In the case of the coefficients $D_{q,i}^{c*}$, they can be written as
\cite{GHD07}
\begin{equation}
\label{2.15bis}
D_{q,1}^{c*}=D_{q,11}^{c*}+D_{q,21}^{c*}, \quad D_{q,2}^{c*}=D_{q,12}^{c*}+D_{q,22}^{c*},
\end{equation}
where the coefficients $D_{q,ij}^{c*}$ have the explicit forms
\begin{eqnarray}
\label{2.18}
D_{q,ij}^{c*}&=&\frac{3}{2}\frac{\pi^{d/2}}{d(d+2)\Gamma \left(\frac{d}{2}\right)}n\sigma_2^d\sum_{p=1}^2
x_p(\sigma_{ip}/\sigma_{12})^d\chi_{ip}M_{ip}(1+\alpha_{ip})
\Big\{\left[(5-\alpha_{ij})M_{ip}-(1-\alpha_{ij})M_{pi}\right]d_{q,pj}^*
\nonumber\\
& &  +(m_1+m_2)x_jD_{pj}^*\left[\frac{\gamma_p}{m_p}\left((5-\alpha_{ip})M_{ip}-(1-\alpha_{ip})M_{pi}\right)
\right.\nonumber\\
& & \left.
+\frac{\gamma_i}{m_i}\left((3+\alpha_{ip})M_{pi}
-(7+\alpha_{ip})M_{ip}\right)\right]
-\frac{32}{3\sqrt{\pi}}\frac{x_pm_p}{m_1+m_2}(\sigma_{12}/\sigma_2)^{d}(\sigma_{ip}/\sigma_{12})
C_{ipj}^*\Big\},
\end{eqnarray}
where
\begin{eqnarray}
C_{ipj}^*&=&
(\theta_i+\theta_p)^{-1/2}(\theta_i\theta_p)^{-3/2}\left\{
\delta_{jp}\beta_{ip}(\theta_i+\theta_p)-\frac{1}{2}\theta_i\theta_p \left[1+\frac{M_{pi}(\theta_i+\theta_p)-2 \beta_{ip}}{\theta_p}\right]
\frac{\partial \ln \gamma_p}{\partial \ln
n_j}\right\}\nonumber\\
& & +\frac{1}{4} (1-\alpha_{ip})(M_{pi}-M_{ip})
\left(\frac{\theta_i+\theta_p}{\theta_i\theta_p}\right)^{3/2} \left( \delta_{jp}
+\frac{3}{2}\frac{\theta_i}{\theta_i+\theta_p}\frac{\partial \ln \gamma_p}{\partial \ln
n_j}\right).\nonumber\\
\label{2.20}
\end{eqnarray}

\subsection{Pressure tensor}
\label{subB}

The overall constitutive relation for the pressure tensor is a combination
of the zeroth $({\bf P}^{(0)})$ and first-order
$({\bf P}^{(1)})$ contributions, which is given by
\begin{equation}
{\bf P}={\bf P}^{(0)}+{\bf P}^{(1)}
\end{equation}
where
\begin{equation}
P_{\alpha\beta}^{(0)}=p \delta_{\alpha\beta}.
\end{equation}
The zeroth-order contribution to the pressure tensor is proportional to the
mixture granular pressure $p$.  The equation of state that defines $p$ is
given by
\begin{equation}
p=p^k+p^c
\end{equation}
where the kinetic $(p^k)$ and the collisional $(p^c)$ contributions are \cite{GDH07}
\begin{equation}
\label{pc}
p^k=nT\quad,\quad p^c=\frac{\pi^{d/2}}{d\Gamma \left(\frac{d}{2}\right)}p^k\;n\sigma_2^d\sum_{i=1}^2\sum_{j=1}^2\;x_ix_j(\sigma_{ij}/\sigma_2)^d
M_{ji}\left(1+\alpha_{ij}\right)\chi_{ij}\gamma_i.
\end{equation}

The constitutive equation for the pressure tensor $P_{\alpha\beta}^{(1)}$,
proportional to the velocity gradients, is
\begin{equation}
P_{\alpha\beta}^{(1)}=-\eta\left(\nabla_\alpha U_\beta+
\nabla_\beta U_\alpha
-\frac{2}{d} \nabla\cdot{\bf U}\delta_{\alpha\beta}\right)
-\kappa \nabla\cdot{\bf U}\delta_{\alpha\beta}.
\label{4.1}
\end{equation}
Here, $\eta $ is the shear viscosity and $\kappa $ is the bulk viscosity. The coefficient $\eta $
has kinetic and collisional contributions while $\kappa $ only has a collisional contribution $\kappa ^{c}$
(and so, vanishes for dilute gases)
\begin{equation}
\eta =\eta ^{k}+\eta ^{c},\quad \kappa =\kappa ^{c}.  \label{3.1a}
\end{equation}

The kinetic part $\eta ^{k}$ is
\begin{equation}
\label{4.1.0}
\eta ^{k}=\eta_1^{k}+\eta_2^{k},
\end{equation}
where the partial contributions $\eta_i^{k}$ can be written as
\begin{equation}
\label{4.1.1}
\eta_i^{k}=\frac{nT}{\nu_0}\eta_i^{k*}.
\end{equation}
The reduced coefficients $\eta_{i}^{k*}$ are given by
\begin{equation}
\label{4.6}
\eta_{1}^{k*}=\frac{2(2\tau_{22}^*-2\zeta^*)\overline{\eta}_1-4\tau_{12}^*\overline{\eta}_2}
{\zeta^{*2}-2\zeta^*(\tau_{11}^*+\tau_{22}^*)+4(\tau_{11}^*\tau_{22}^*-\tau_{12}^*\tau_{21}^*)},
\end{equation}
\begin{equation}
\label{4.6.0}
\eta_{2}^{k*}=\frac{2(2\tau_{11}^*-2\zeta^*)\overline{\eta}_2-4\tau_{21}^*\overline{\eta}_1}
{\zeta^{*2}-2\zeta^*(\tau_{11}^*+\tau_{22}^*)+4(\tau_{11}^*\tau_{22}^*-\tau_{12}^*\tau_{21}^*)},
\end{equation}
where the expressions of the (reduced) collision frequencies $\tau_{ij}^*$
can be found in the Appendix A of Ref.\ \cite{GHD07}.
In Eq.\ (\ref{4.6}), we have introduced the quantities
\begin{equation}
\label{4.7}
\overline{\eta}_1=x_1\gamma_1+E_{11}+E_{12}, \quad \overline{\eta}_2=x_2\gamma_2+E_{22}+E_{21},
\end{equation}
where
\begin{eqnarray}
E_{ij}&=&\frac{\pi ^{d/2}}{d(d+2)\Gamma \left( \frac{d}{2} \right) }n\sigma_2^d
x_ix_j
(\sigma_{ij}/\sigma_2)^d\chi_{ij}M_{ji}(1+\alpha _{ij})\nonumber\\
& & \times \left[
M_{ji}(3\alpha _{ij}-1)\left( \gamma_i+\frac{m_i}{m_{j}}\gamma_{j}\right)
-4M_{ij}(\gamma_{i}-\gamma_{j})\right].  \label{3.7}
\end{eqnarray}

The collisional contribution $\eta ^{c}$ to the shear viscosity and the the bulk viscosity $\kappa$ have the forms
\begin{equation}
\label{4.4}
\eta^{c}=\frac{nT}{\nu_0}\eta ^{c*}, \quad \kappa=\frac{nT}{\nu_0}\kappa^*,
\end{equation}
where
\begin{equation}
\eta^{c*}=\frac{2\pi ^{d/2}}{\Gamma \left( \frac{d}{2}\right)}
\frac{1}{d(d+2)}n\sigma_2^d\sum_{i=1}^{2}\sum_{j=1}^{2}x_{j}
(\sigma_{ij}/\sigma_2)^d\chi_{ij}M_{ji}
(1+\alpha _{ij})\eta_{i}^{k*}+\frac{d}{d+2}\kappa^{*},  \label{4.5}
\end{equation}
\begin{equation}
\label{4.3}
\kappa^{*}=\frac{4\pi ^{(d-1)/2}}{d^2\Gamma \left( \frac{d}{2}\right)}\frac{(n\sigma_2^d)^2}{m_{1}+m_{2}}
\sum_{i=1}^{2}\sum_{j=1}^{2} x_{i}x_{j}\frac{m_im_{j}}{m_i+m_j}
\frac{\sigma_{12}^{d-1}\sigma_{ij}^{d+1}}{\sigma_2^{2d}}
\chi_{ij}(1+\alpha _{ij})\left( \frac{\theta _{i}+\theta _{j}}{\theta_{i}\theta_{j}}\right)^{1/2}.
\end{equation}

It must be remarked that the predictions of the shear viscosity $\eta$ compare quite well with Monte Carlo simulations of a heated granular binary mixture, even for strong dissipation \cite{GS03}.

\subsection{Cooling rate}
\label{subC}

The overall cooling rate can be written as the sum of the zeroth-order
$(\zeta^{(0)})$ and first-order contributions $(\zeta_u)$
\begin{equation}
\label{4.8}
\zeta=\zeta^{(0)}+\zeta_u\nabla\cdot{\bf U}.
\end{equation}
The zeroth-order cooling rate of each species $(\zeta_i^{(0)})$ defines
the rate of kinetic energy loss for that species, and is given by the
following relation
\begin{eqnarray}
\label{4.9}
\zeta^{(0)}=\zeta_i^{(0)}&=&\frac{4\pi^{(d-1)/2}}{d\Gamma \left(\frac{d}{2}\right)}\nu_0\sum_{j=1}^2\chi_{ij}x_jM_{ji}(\sigma_{ij}/\sigma_{12})^{d-1}
\left(\frac{\theta_i+\theta_j}{\theta_i\theta_j}\right)^{1/2}
\left(1+\alpha_{ij}\right)\nonumber\\
& & \times \left[1-\frac{M_{ji}}{2}\left(1+\alpha_{ij}
\right)\frac{\theta_i+\theta_j}{\theta_j}\right].
\end{eqnarray}
As shown in Eq.\ (\ref{4.9}), the zeroth-order cooling rate for each species is
equivalent (i.e., $\zeta_1^{(0)}=\zeta_2^{(0)}$).  Because
Eq.\ (\ref{4.9}) is an implicit expression that depends on individual species
granular temperatures, the following equation is needed
\begin{equation}
\label{4.10}
nT=n_1T_1+n_2T_2.
\end{equation}
The expressions given in Eq.\ (\ref{4.9}) and Eq.\ (\ref{4.10}) form a set of two non-linear
algebraic equations that can be solved for $\theta_1$ and $\theta_2$
(using the relation $\theta_i=m_iT/mT_i$), and then species temperatures $T_1$ and $T_2$ can subsequently be found.  The equation of state defining
$\zeta^{(0)}$ was first proposed by Garz\'{o} and Dufty \cite{GD99}.

At first order in gradients, there is a contribution to the cooling rate
from $\nabla \cdot \mathbf{U}$. The proportionality coefficient $\zeta _{u}$
is a new transport coefficient for granular fluids. Two different
contributions can be identified
\begin{equation}
\zeta _{u}=\zeta ^{(1,0)}+\zeta ^{(1,1)}.  \label{5.1}
\end{equation}
The coefficient $\zeta ^{(1,0)}$ is given by
\begin{equation}
\zeta ^{(1,0)}=-\frac{3\pi ^{d/2}}{d^{2}\Gamma \left( \frac{d}{2} \right)
}n\sigma_2^d\sum_{i=1}^{2}\sum_{j=1}^{2}x_{i}x_{j}M_{ji}(\sigma_{ij}/\sigma_2)^d
\chi _{ij}(1-\alpha _{ij}^{2})\gamma_i.
\label{5.2}
\end{equation}

The contribution $\zeta ^{(1,1)}$ can be written as
\begin{equation}
\zeta ^{(1,1)}=\frac{3\pi ^{(d-1)/2}}{d\Gamma \left( \frac{d}{2} \right)
}\sum_{i=1}^{2}\sum_{j=1}^{2}\frac{m_j}{m_1+m_2}x_ix_j
(\sigma _{ij}/\sigma_{12})^{d-1}\chi_{ij}M_{ij}(1-\alpha
_{ij}^{2})\theta _{i}^{-3/2}\theta _{j}^{1/2}(\theta _{i}+\theta _{j})^{-1/2}e_{i,D}^*,  \label{5.3}
\end{equation}
where
\begin{equation}
\label{5.4}
e_{1,D}^{*}=\frac{2(2\psi_{22}^*-3\zeta^*)\overline{e}_{1,D}-4\psi_{12}^*\overline{e}_{2,D}}
{9\zeta^{*2}-6\zeta^*(\psi_{11}^*+\psi_{22}^*)+4(\psi_{11}^*\psi_{22}^*-\psi_{12}^*\psi_{21}^*)},
\end{equation}
\begin{equation}
\label{5.4.1}
e_{2,D}^{*}=\frac{2(2\psi_{11}^*-3\zeta^*)\overline{e}_{2,D}-4\psi_{21}^*\overline{e}_{1,D}}
{9\zeta^{*2}-6\zeta^*(\psi_{11}^*+\psi_{22}^*)+4(\psi_{11}^*\psi_{22}^*-\psi_{12}^*\psi_{21}^*)}.
\end{equation}
Here, the collision frequencies $\psi_{ij}^*$ have been determined in the
Appendix A of Ref.\ \cite{GHD07} and the coefficients $\overline{e}_{i,D}$ are given by
\begin{eqnarray}
\overline{e}_{i,D} &=&-\frac{\pi^{d/2}}{2d^2(d+2)
\Gamma \left(\frac{d}{2}\right)} n\sigma_2^d\sum_{j=1}^{2}x_{j}(\sigma_{ij}/\sigma_{12})^{d}
\chi _{ij}M_{ji}(1+\alpha_{ij})
\left[ 8(d+2)(M_{ij}-1)\right.  \notag \\
&&+4(13+2d+9\alpha _{ij})M_{ji}-48 M_{ji}^{2}\theta_{j}^{-1}(\theta
_{i}+\theta _{j})(1+\alpha_{ij})^{2}  \notag \\
&&\left. +15 M_{ji}^{3}\theta_{j}^{-2}(\theta_{i}+\theta
_{j})^{2}(1+\alpha_{ij})^{3}\right] .  \label{5.5}
\end{eqnarray}

The results displayed along this section give the explicit forms for the equations of
state, the transport coefficients and the cooling rate of a moderately dense granular
binary mixture. The corresponding expressions for a low-density binary mixture can be easily obtained from their dense forms by taking the limit $n\sigma_2^d\to 0$. These explicit expressions are displayed in the Appendix \ref{appC} and agree with those previously derived from the Boltzmann equation \cite{GD02,GM07}.

\section{Quantitative approach: comparison of dilute and dense-phase
expressions for hard spheres}
\label{sec3}

In order to assess the importance of dense-phase corrections to the
continuum theory for rapid granular flows of binary mixtures, the
equations of state and transport coefficients obtained from the GHD and GD
theories were compared over a range of volume fractions and coefficients of restitution
for a given set of mixture properties (diameter ratio, size
ratio, and volume fraction ratio). To illustrate the differences in a
straightforward manner, each quantity is examined as a ratio of the GHD
value (dilute through moderately dense) to the GD value (dilute limit),
giving rise to a non-dimensional quantity.  These non-dimensional ratios
were plotted as functions of volume fraction and coefficients of restitution,
holding all other mixture properties constant. Representing the transport
coefficients and equations of state in this manner reveals the relative
magnitudes of the dense- and dilute-phase predictions. Recall the complete
set of equations of state and transport coefficients for GHD theory are
given in Table 1 ($\zeta^{(0)}$, $\zeta_u$, $D_{ij}$, $D_i^T$, $D_{ij}^F$,
$\lambda$, $D_{q,i}$, $L_{ij}$, $p$, $\eta$, $\kappa$).

It is important to note
that some transport coefficients ($L_{ij}$, $D_{ij}^F$) were not
considered in the dilute theory (GD), and thus these quantities are
not considered here. Moreover, two of the transport
coefficients, namely $\zeta_u$ and $\kappa$, are zero in the dilute limit,
and thus the corresponding
ratios of the moderately dense (GHD) value to the dilute (GD) value
diverge.  Accordingly, only the GHD predictions of these quantities are
shown.  Thus, the comparison between
the GHD and GD theory predictions presented here involves the seven
remaining quantities: $\zeta^{(0)}$, $p$, $\eta$, $D_{ij}$, $D_i^T$,
$\lambda$, $D_{q,i}$.

\subsection{Mixture parameters}

The continuum description of a binary mixture of inelastic hard spheres ($d=3$) is a function of the following dimensional parameters: species masses ($m_1$, $m_2$), species
diameters ($\sigma_1$, $\sigma_2$), species 1 volume fraction ($\phi_1$),
overall volume fraction ($\phi=\phi_1+\phi_2$), and coefficients of
restitution ($\alpha_{11},\alpha_{22},\alpha_{12}=\alpha_{21}$).
(Note that the number densities and volume fractions are related by
$\phi_i=n_i\pi\sigma_i^3/6$.)  The subscripts 1
and 2 denote the two species in the binary mixture. For purposes of simplicity, the coefficients of restitution have been assumed to be the same for all
combinations of collisions (i.e., $\alpha_{11}=\alpha_{22}=\alpha_{12}\equiv \alpha$). In terms of the ratio of moderately dense
(GHD) to dilute (GD) predictions for each quantity, the parameter space is
reduced to the following dimensionless inputs: mass ratio ($m_1/m_2$),
diameter ratio ($\sigma_1/\sigma_2$), overall volume fraction ($\phi$),
volume fraction ratio of species 1 relative to the total ($\phi_1/\phi$),
and coefficient of restitution $\alpha$.
Hereafter, the ratio $\phi_1/\phi$
will be referred to as the (mixture) composition of species 1. Recall that
the GHD and GD theories allow for a non-equipartition of energy, and
thus several of the aforementioned closures
(see, for example, Eq.\ (\ref{2.15})) involve the species granular temperatures,
$T_1$ and $T_2$.  It is important to point out that these quantities are not
hydrodynamic variables (i.e., they do not require the solution of species
energy balances; for a detailed explanation, see Ref.\ \cite{GHD07}) and instead are determined by the set of equations
defining the zeroth-order cooling rate (Eqs.\ (\ref{4.8}) and (\ref{4.9})).

\subsection{Parameter space evaluated}

Table 2 summarizes the three cases (equal size and different mass, equal
mass and different size, and different size and mass) used to compare the
GHD and GD theories, and the wide ranges of input parameters used in each
case study. Though the transport coefficients and equations of state may
vary quantitatively from case to case, the general trends show little
variation. For sake of brevity, the upcoming section will focus on one
representative case, namely that of different size and equal material
densities (i.e., different mass) in order to quantify how the newly
acquired GHD predictions differ from the dilute-phase counterpart (GD).

\begin{table}[htb]
\begin{center}
{\bf Table 2:} Range of input parameters used in analysis of binary mixture
via GHD theory\smallskip

\begin{tabular}{|c|c|c|c|c|}\hline
{\bf Case Description} & \ \ \ \ $\sigma_1/\sigma_2$ & \ \ \ \ $m_1/m_2$ &
$\phi_1/\phi$ & $\alpha$\\ \hline
Equal Sizes & 1 & 1-10 & \ \ 0.25-0.75 & \ \ 0.50-0.99\\ \hline
Equal Masses & 1-10 & 1 & 0.25-0.75 & 0.50-0.99\\ \hline
\ \ Different diameters, different masses
& 2 & \ \ 0.10-10 & 0.50 & 0.75\\ \hline
\end{tabular}\end{center}\end{table}\vspace*{-.4in}

\subsection{Case presented: Different-sized particles with equal material
densities}

Many industrial and natural granular flows are comprised of one material
(i.e., same material density), but different-sized particles.  In the case
presented here, the diameter of species 1 was twice that of species 2
(i.e., $\sigma_1/\sigma_2=2$),  and both species had the same material density
(i.e., $m_1/m_2=8$). For the sake of consistency, the composition of each species was
held constant at 50\% by volume for this analysis (i.e., $\phi_1/\phi=\phi_2/\phi=0.5$).
The ratio of GHD to GD predictions of each quantity evaluated was
plotted over a range of volume fractions from dilute to moderately dense
($\phi=10^{-8}$-0.5) while holding the coefficient of restitution constant. Also,
each quantity was varied over a range of coefficients of restitution from
relatively inelastic to nearly elastic (0.5-0.99) while holding the
overall volume fraction constant. The results of this case study are
presented in the upcoming section.

\section{Results and discussion}
\label{sec4}

The overall objective was to analyze each transport coefficient and
equation of state over a range of parameters for the newly-developed
GHD theory. By comparing these results to the predictions from the
dilute (GD) theory, it was possible to demonstrate the need for a
moderately dense-phase correction, as detailed below.

\begin{figure}[htb]
\includegraphics[width=0.4\columnwidth]{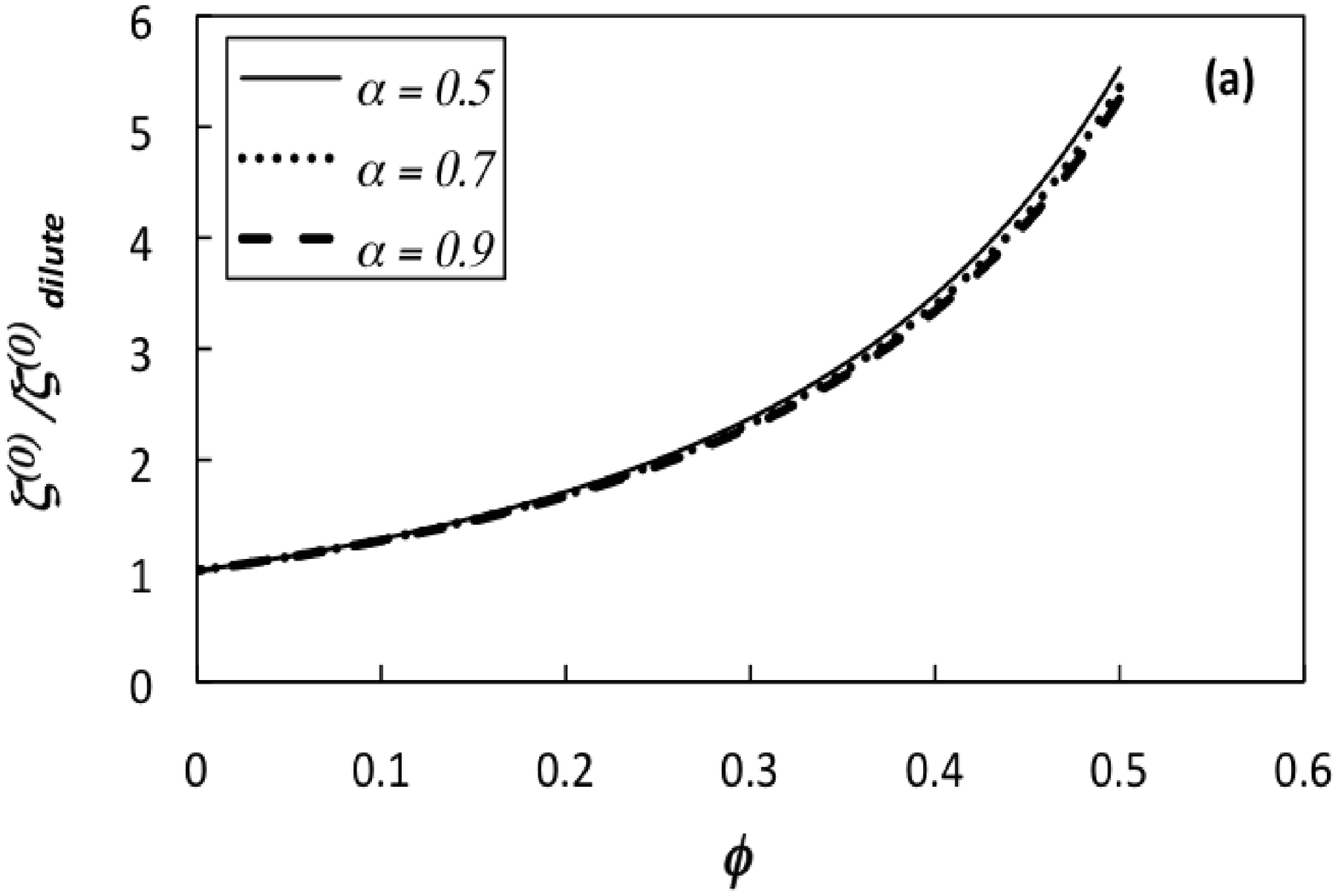}
\includegraphics[width=0.4\columnwidth]{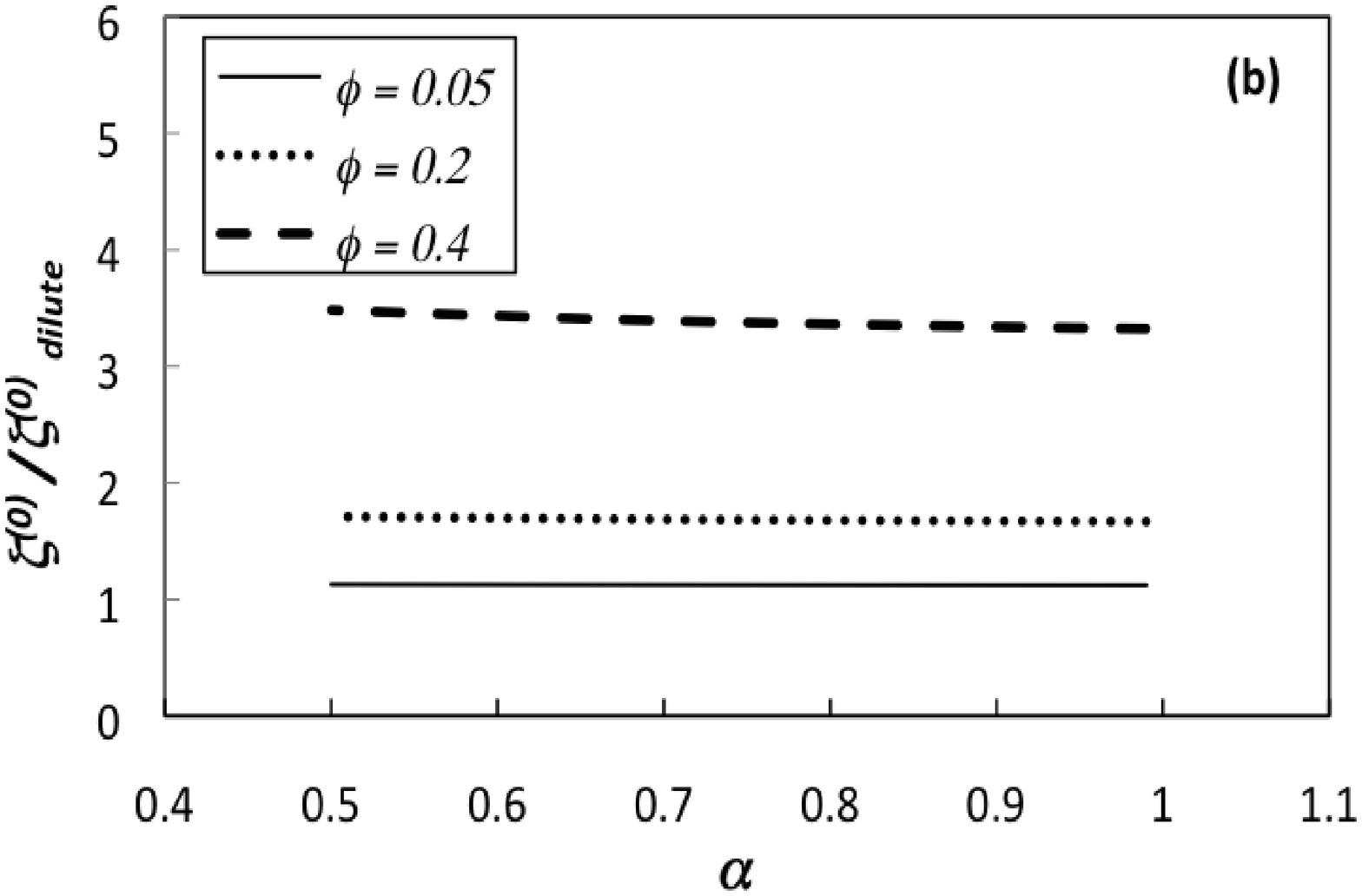}
\caption{Zeroth-order cooling rate:  ratio of moderately dense (GHD)
to dilute (GD) predictions as a function of (a)~overall volume
fraction and (b)~coefficient of restitution.}
\end{figure}
\begin{figure}[htp]
\includegraphics[width=0.4\columnwidth]{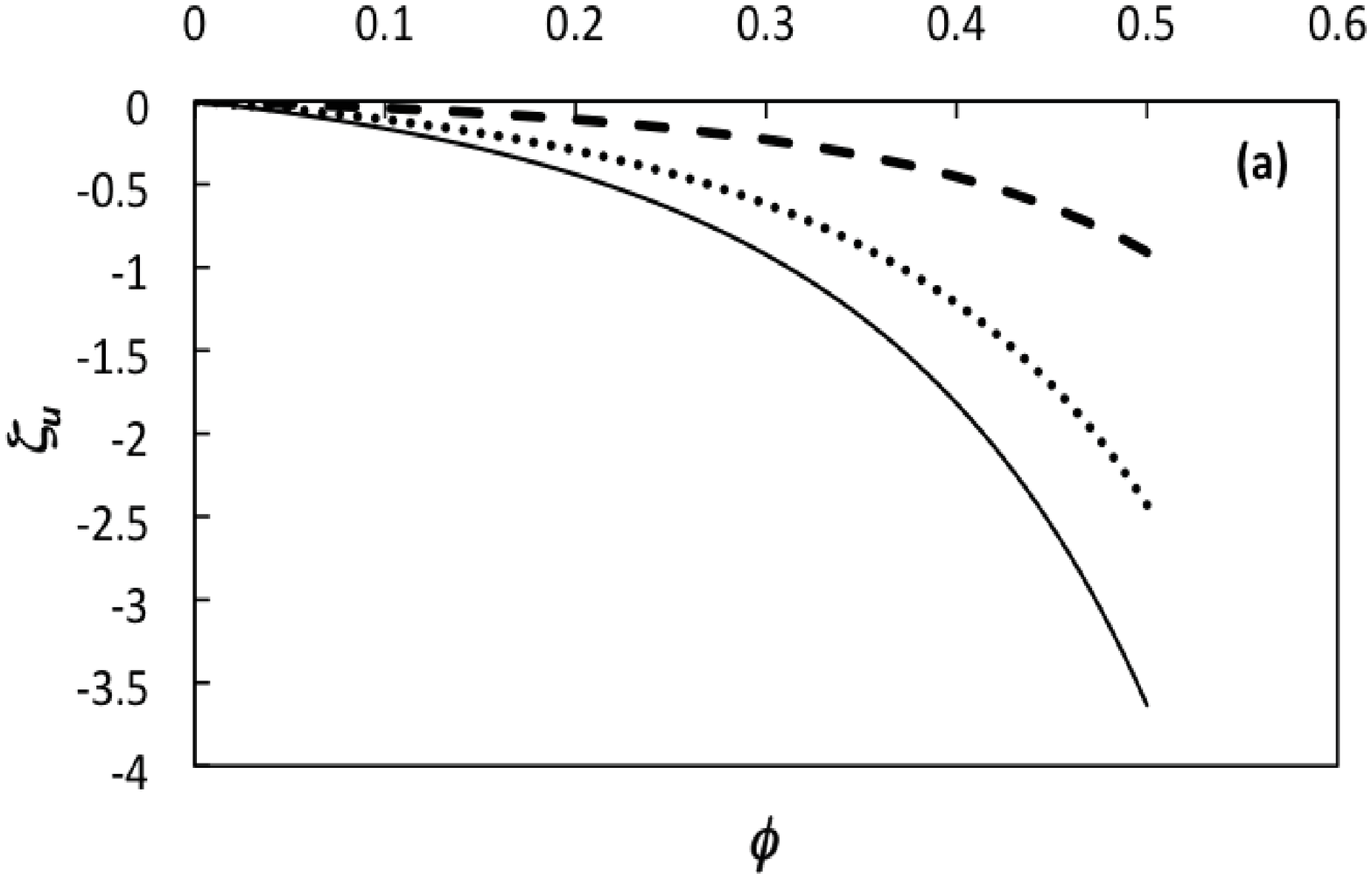}
\includegraphics[width=0.4\columnwidth]{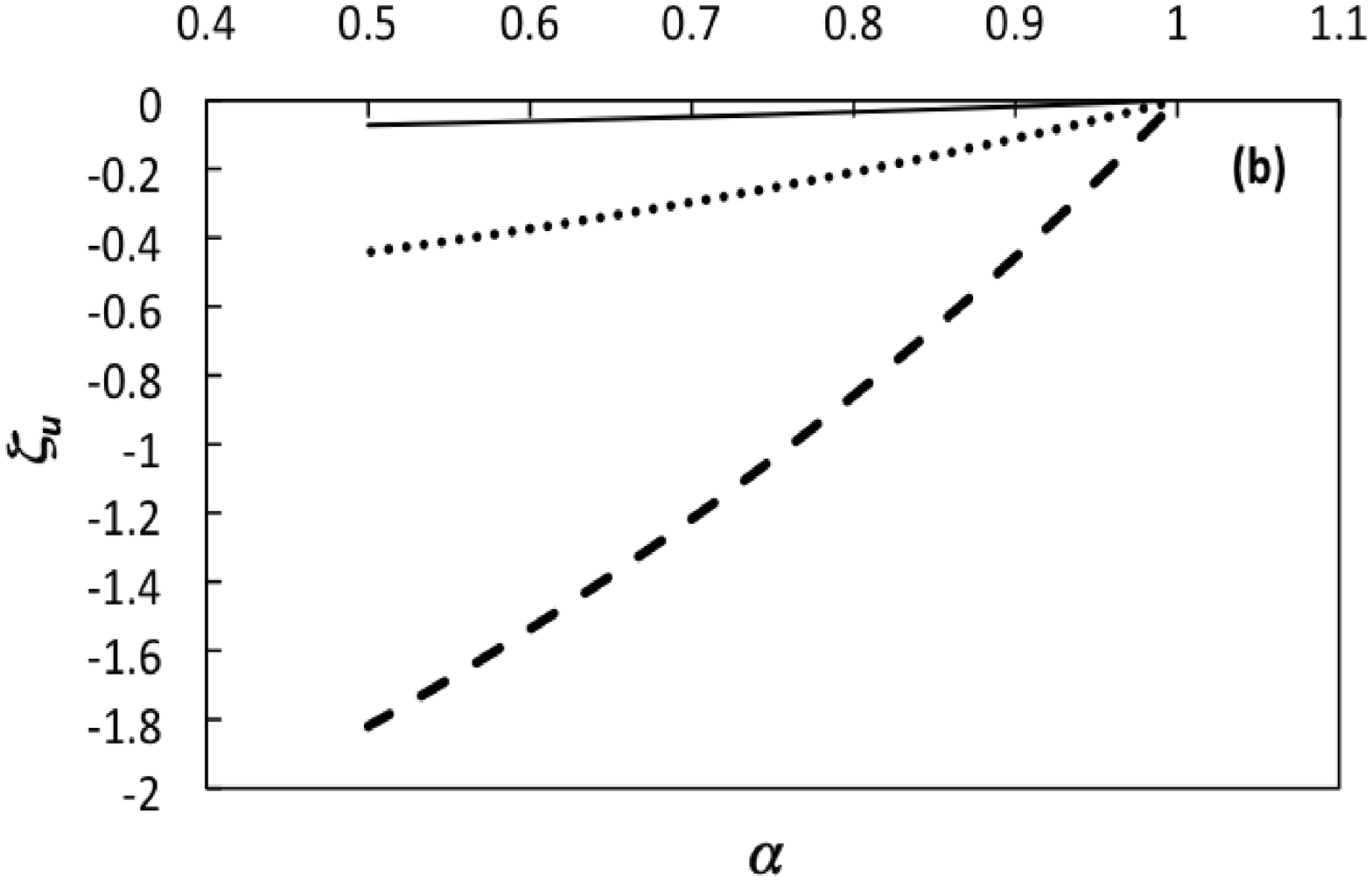}
\caption{
Transport coefficient associated with first-order cooling
rate:  moderately dense (GHD) predictions as a function of
(a)~overall volume fraction and (b)~coefficient of restitution. See legends
presented in Figure 1.}
\end{figure}

\subsection{Cooling rate: zeroth-order and first-order contributions}

As indicated by Figure 1, the dense-to-dilute ratio of the zeroth-order
cooling rate ($\zeta^{(0)}$) is much more sensitive to changes in volume
fraction than coefficient of restitution. Such behavior can be explained
via the dependency of the zeroth-order cooling rate (Eq.\ {\ref{4.9}}) on the pair
correlation function at contact, $\chi_{ij}$ (Eqs.\ (\ref{a15}) and (\ref{a18})). This factor accounts
for the volume exclusion effects between like particles ($\chi_{11}$) and
unlike particles ($\chi_{12}$). In the dilute limit, the spatial correlation
factor equals one (i.e., $\chi_{11}=\chi_{12}=1$). When the zeroth-order
cooling rate of GHD theory is then divided by its dilute counterpart, the
resulting function is strongly dependent on the spatial correlation factor.
Because $\chi_{ij}$ is sensitive to changes in overall volume fraction, it is
then reasonable that the dimensionless zeroth-order cooling rate ratio
exhibits the same sensitivity.  More specifically, the results shown in
Figure 1a indicate that $\zeta^{(0)}$ predicted by GHD theory is more than 5
times greater than its dilute counterpart for a fairly dense system
($\phi=0.5$) and more than 2 times greater for $\phi=0.3$. Even at
$\phi=0.2$, a discrepancy of 27\% is found between the dilute- and dense-phase
predictions.

Unlike the zeroth-order contribution to the cooling rate, the
transport coefficient associated with the first-order contribution is
zero in the dilute limit. Therefore, a ratio comparison of the
dense-to-dilute predictions is not possible. The results given in
Figure 2 represent the first-order contribution to
the cooling rate (which is non-dimensional), which approaches zero as volume fraction
diminishes. As evident from this figure, $\zeta_u$ is quite sensitive to
changes in both volume fraction (Fig.\ 2a) and coefficient of restitution
(Fig.\ 2b). Also, the results for this case indicate that the
magnitude of the first-order contribution increases
as the system becomes denser and less elastic.
\begin{figure}[htp]
\includegraphics[width=0.4\columnwidth]{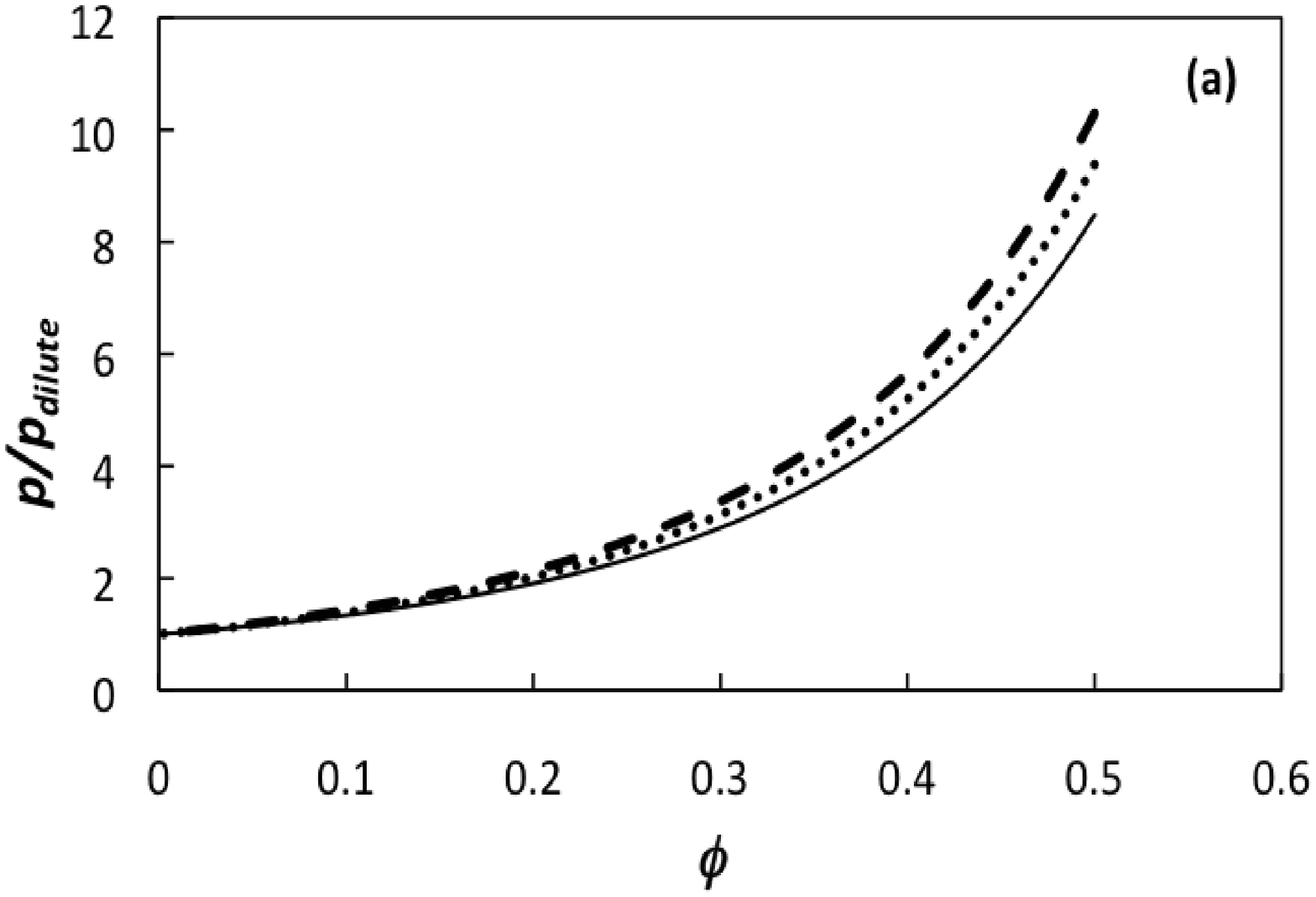}
\includegraphics[width=0.4\columnwidth]{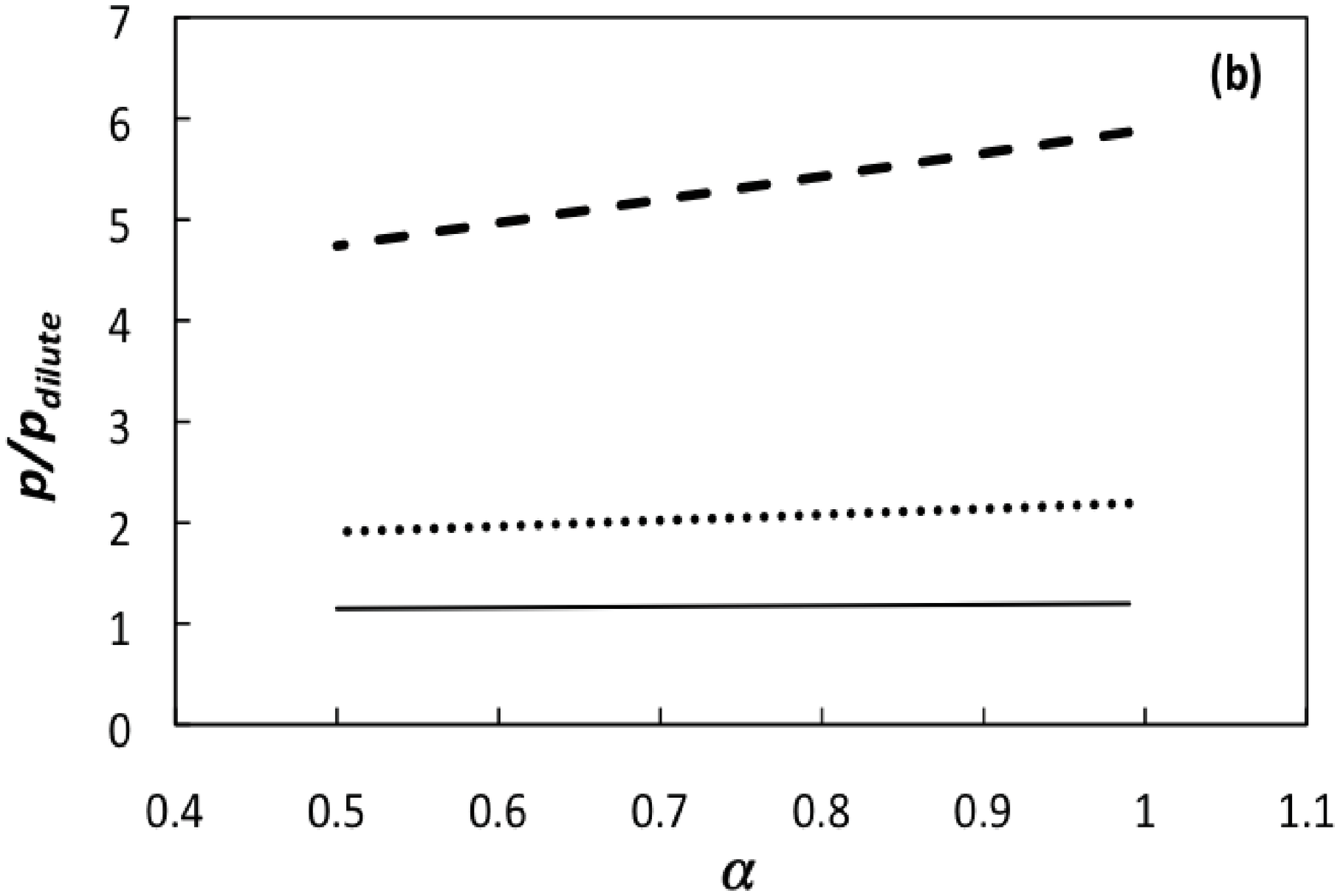}
\caption{
Pressure:  ratio of moderately dense (GHD) to dilute (GD)
predictions as a function of (a)~overall volume fraction and
(b)~coefficient of restitution. See legends presented in Figure 1.
}
\end{figure}
\begin{figure}[htp]
\includegraphics[width=0.4\columnwidth]{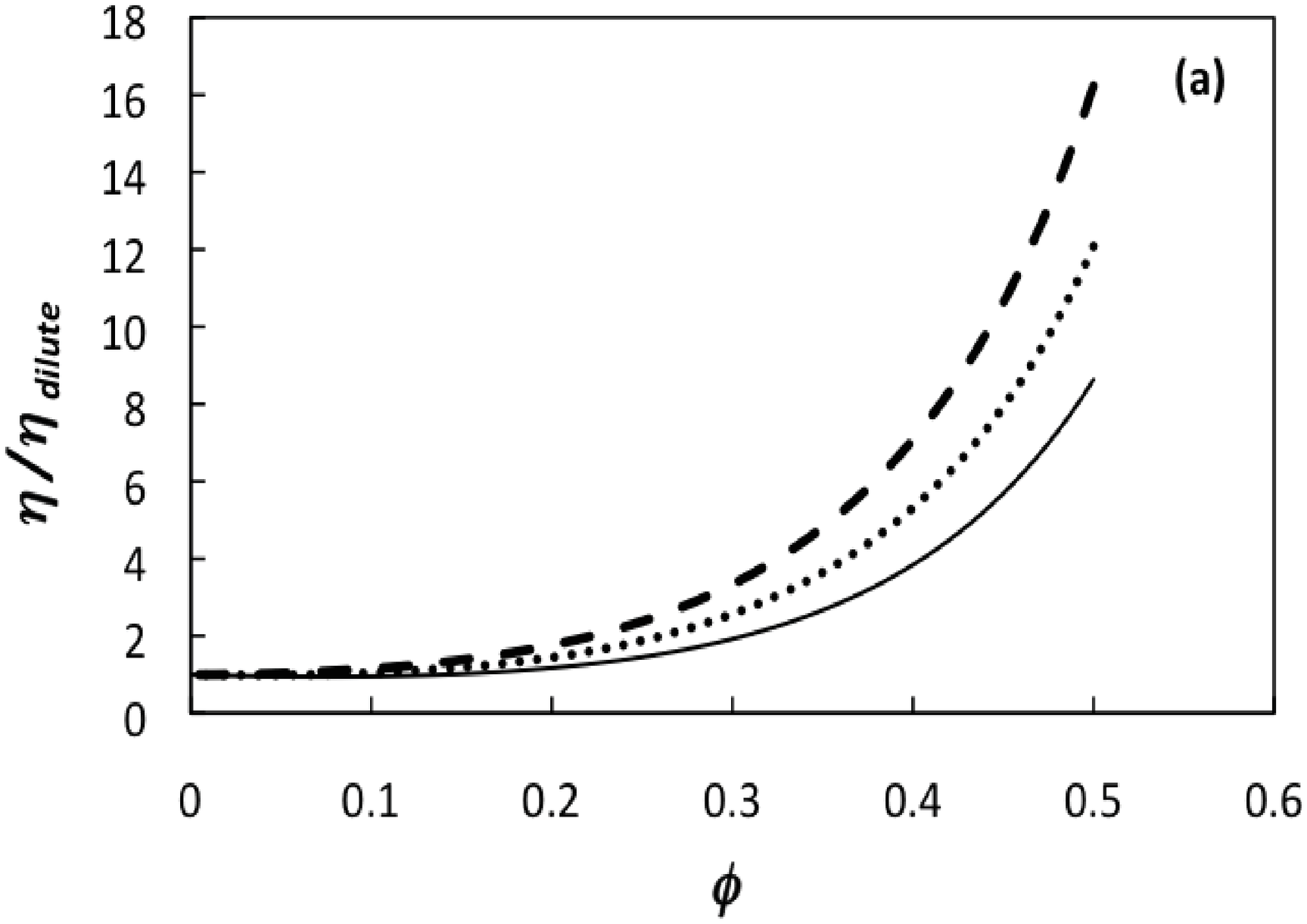}
\includegraphics[width=0.4\columnwidth]{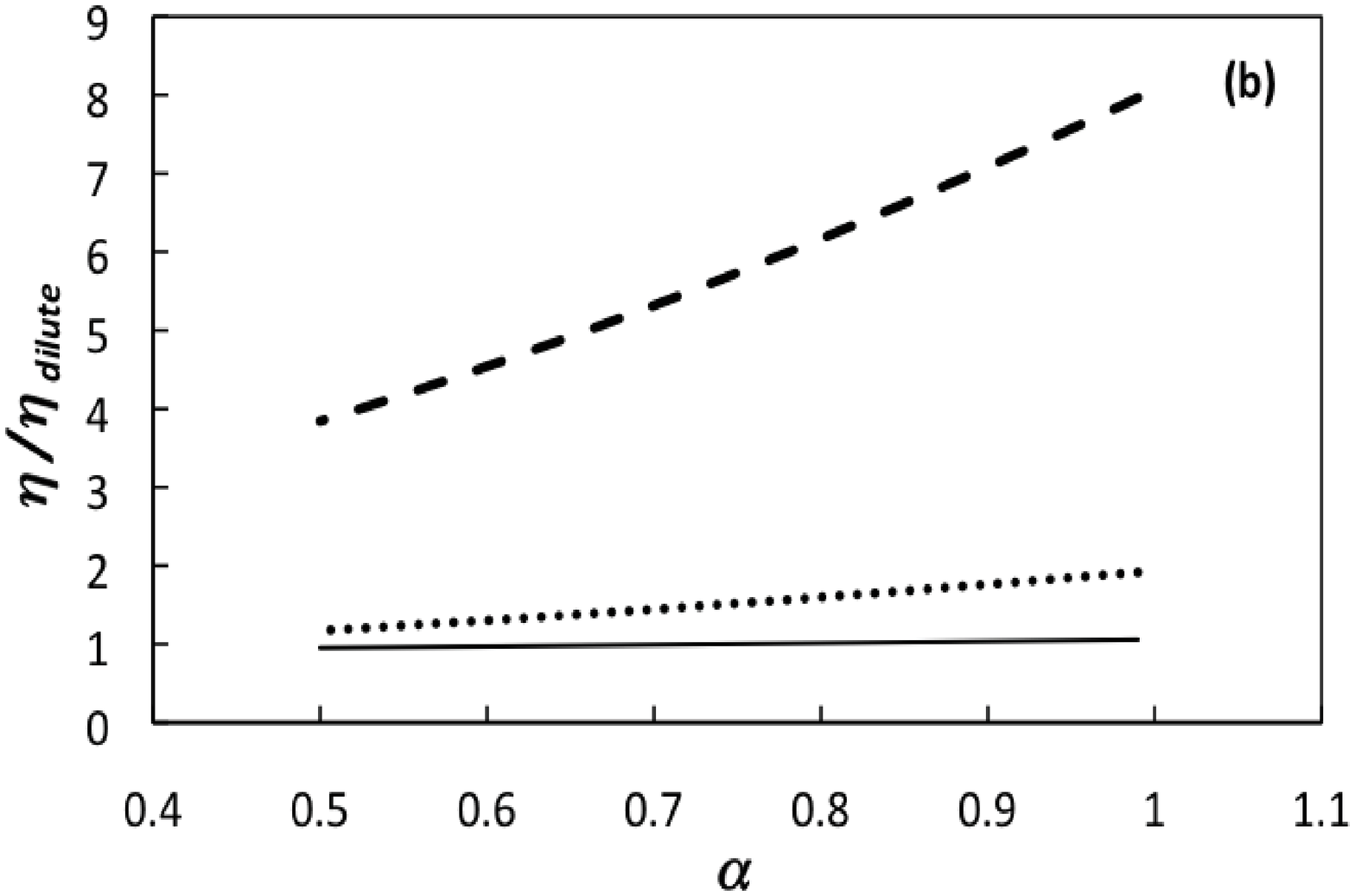}
\caption{
Shear viscosity:  ratio of moderately dense (GHD) to dilute
(GD) predictions as a function of (a)~overall volume fraction and
(b)~coefficient of restitution. See legends presented in Figure 1.
}
\end{figure}

\subsection{Momentum flux: pressure, shear viscosity and bulk viscosity}

Now moving on to results associated with the momentum flux, Figure 3
indicates that the dense-to-dilute ratio of granular pressure is more
sensitive to changes in volume fraction (Fig.\ 3a) than
coefficient of restitution (Fig.\ 3b). Also, this ratio increases monotonically with
both volume fraction and coefficient of restitution. For a moderately
dense system ($\phi\sim 0.4$), GHD theory predicts that the granular pressure
is about 5 times greater than dilute (GD) theory. Even at lower
volume fractions ($\phi\sim 0.1$), the moderately dense-phase prediction is
greater than its dilute counterpart by 40\%.

Shear viscosity, results of which are given in Figure 4, behaves in a
similar manner to granular pressure (Fig.\ 3). A monotonic increase is
exhibited with respect to both volume fraction (Fig.\ 4a) and
coefficient of restitution (Fig.\ 4b). The GHD prediction is about 5 times
larger than the GD prediction for moderately dense systems ($\phi\sim 0.4$).
However, the discrepancy at lower volume fractions ($\phi\sim0.1$) decreases
to approximately 5\% (Fig.\ 4a).
\begin{figure}[htp]
\includegraphics[width=0.4\columnwidth]{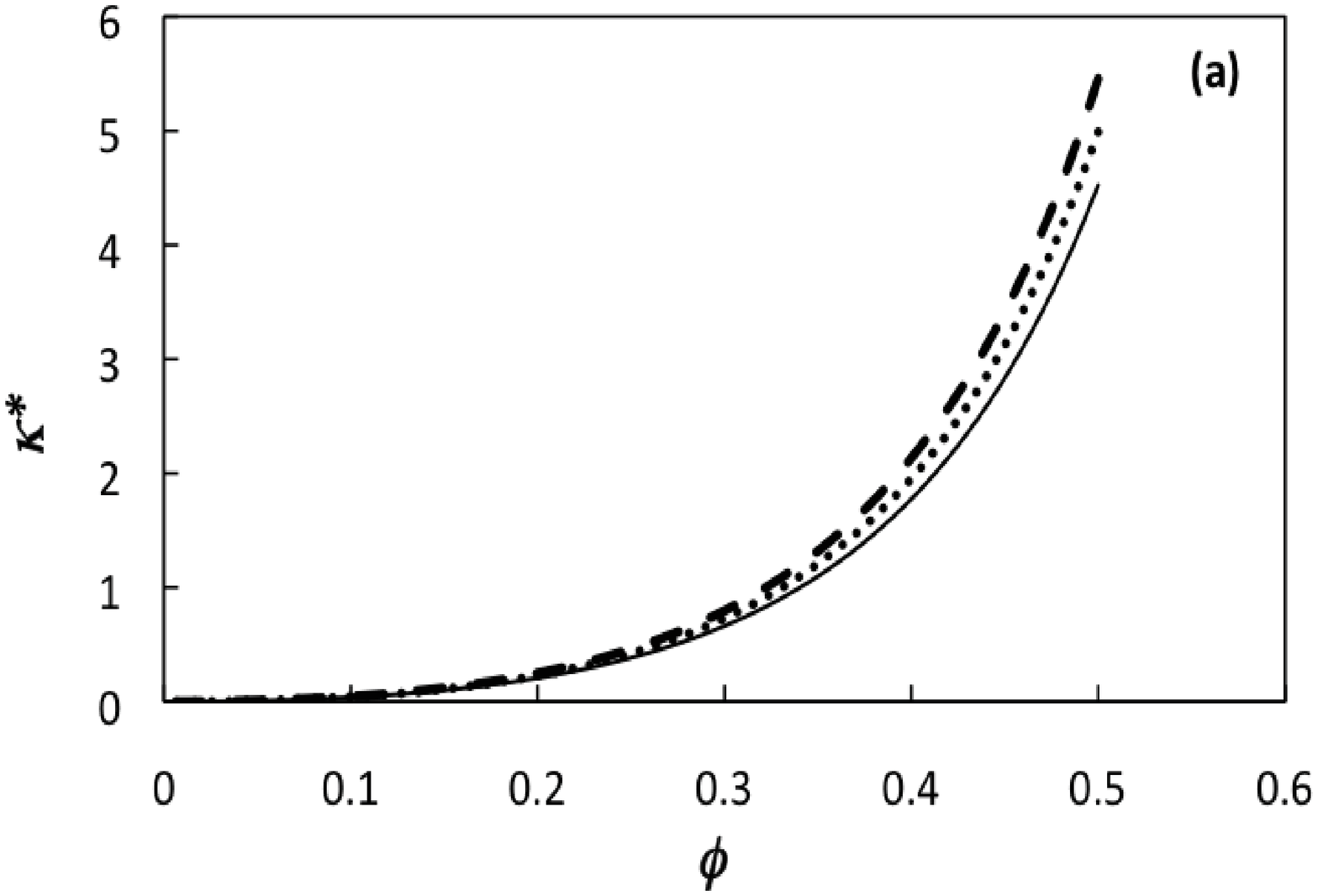}
\includegraphics[width=0.4\columnwidth]{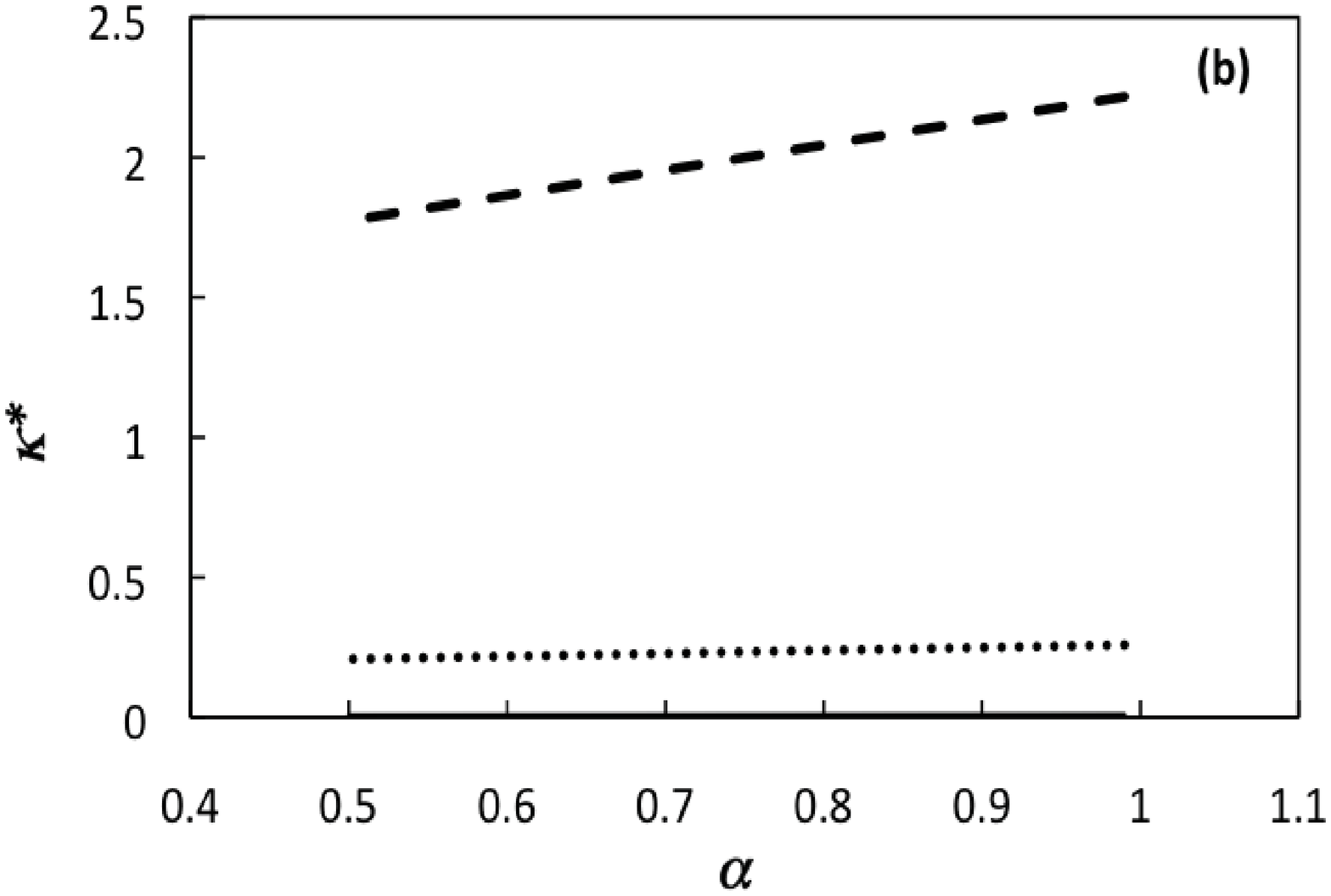}
\caption{
Bulk viscosity (non-dimensional):  moderately dense (GHD) predictions as a function of (a) overall volume
fraction and (b) restitution coefficient. See legends presented in Figure 1. The dimensionless inputs are as follows: $m_1/m_2 = 8$, $\sigma_1/\sigma_2 = 2$,  and $\phi_1/\phi = 0.5$.
}
\end{figure}

As mentioned previously, the bulk viscosity is zero in the dilute limit.
Therefore the results for bulk viscosity, given in Figure 5, are those
obtained from the moderately dense theory (GHD) alone, instead ratios of
dense-to-dilute predictions. Furthermore,
these GHD-based bulk viscosities are non-dimensionalized according to
Eqs.\ (\ref{4.4}) and (\ref{4.3}). It is evident from this figure that
the prediction of bulk viscosity via GHD theory increases significantly
in magnitude as the system becomes moderately dense, whereas little
variation results from changes in particle elasticity.

\begin{figure}[htp]
\includegraphics[width=0.4\columnwidth]{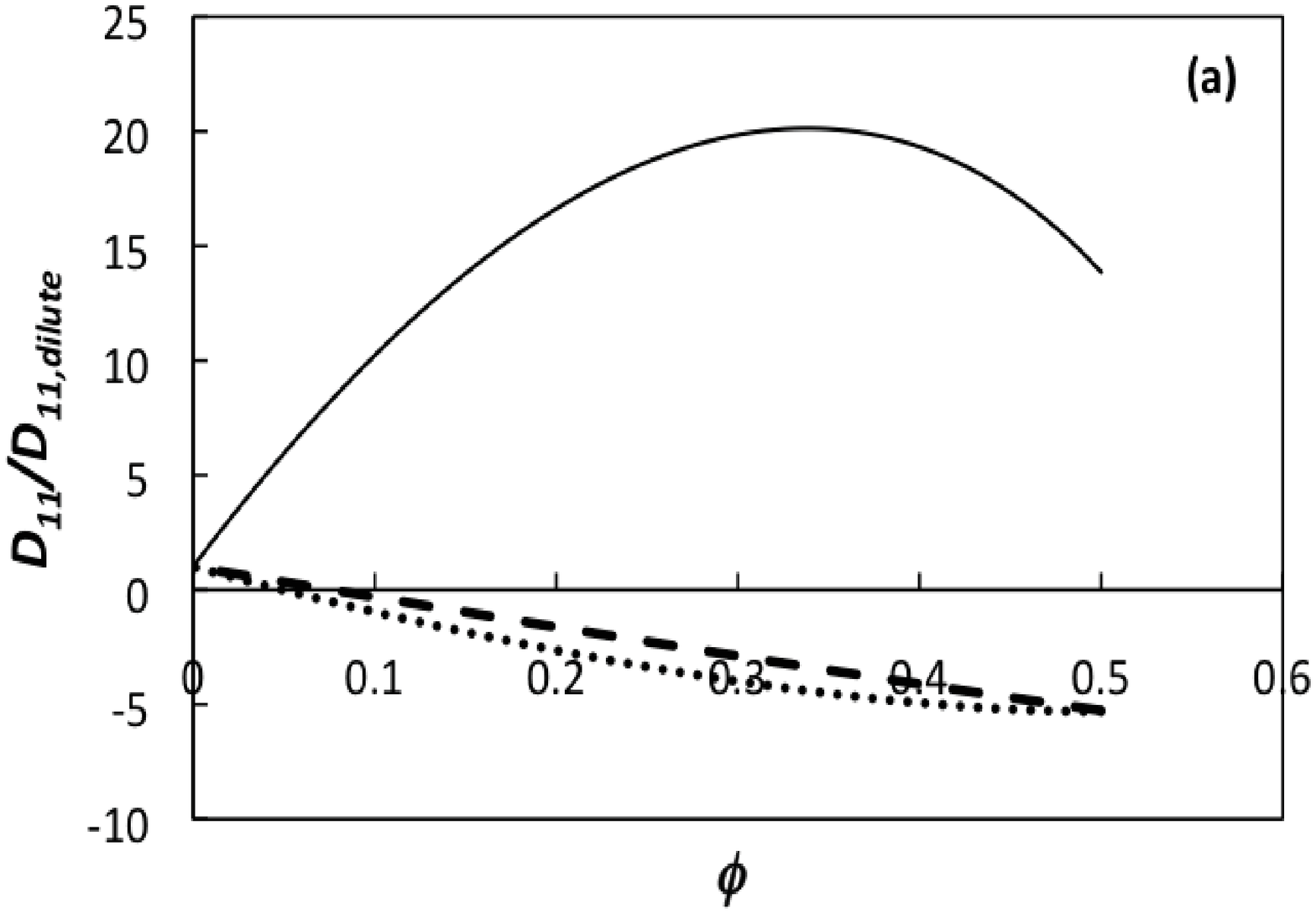}
\includegraphics[width=0.4\columnwidth]{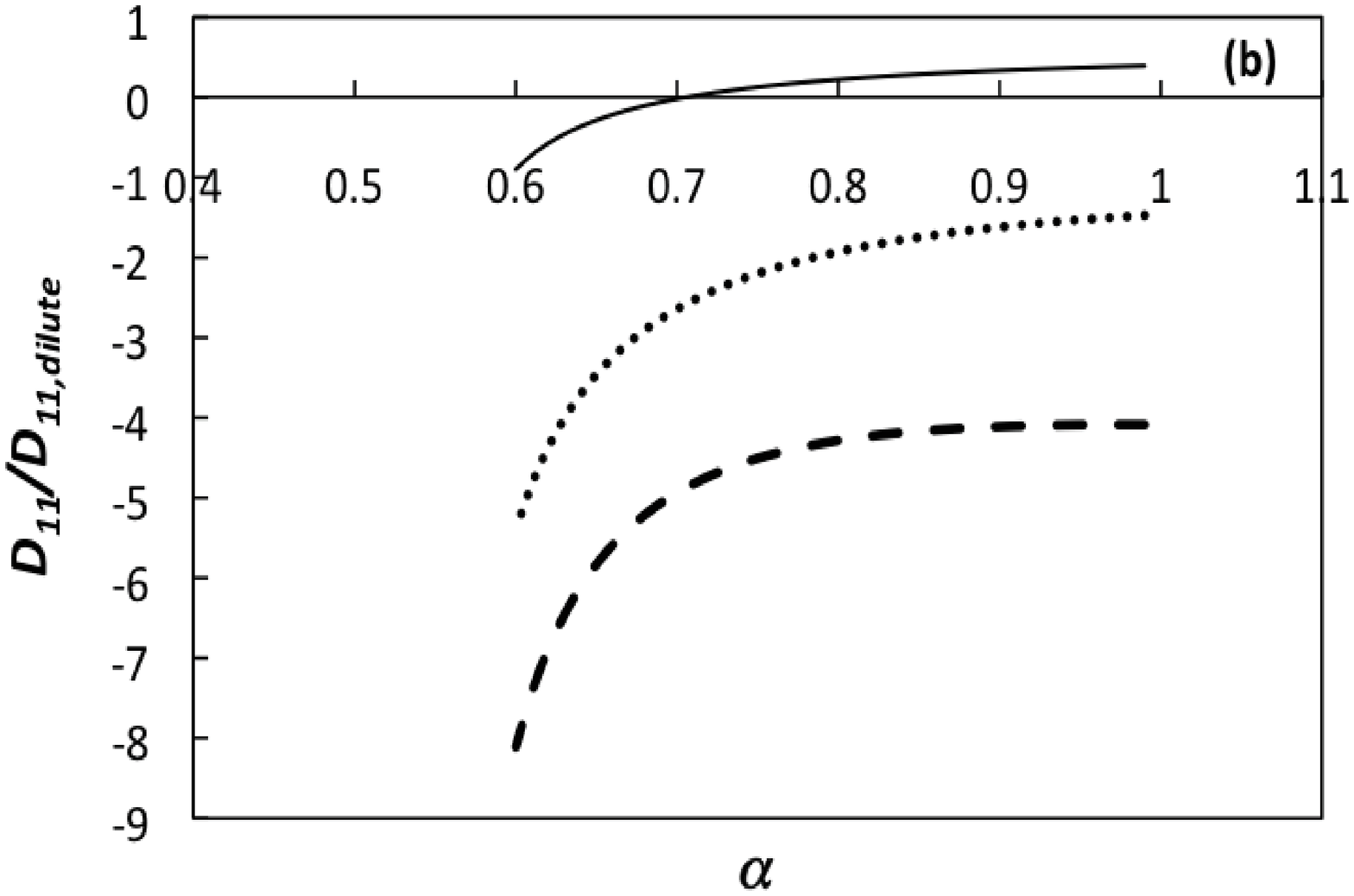}
\caption{
Mutual diffusion ($D_{11}$):  ratio of moderately dense (GHD) to
dilute (GD) predictions as a function of (a)~overall volume fraction
and (b)~coefficient of restitution. See legends presented in Figure 1.
}
\end{figure}
\begin{figure}[htb]
\begin{tabular}{ll}
{\psfig{figure=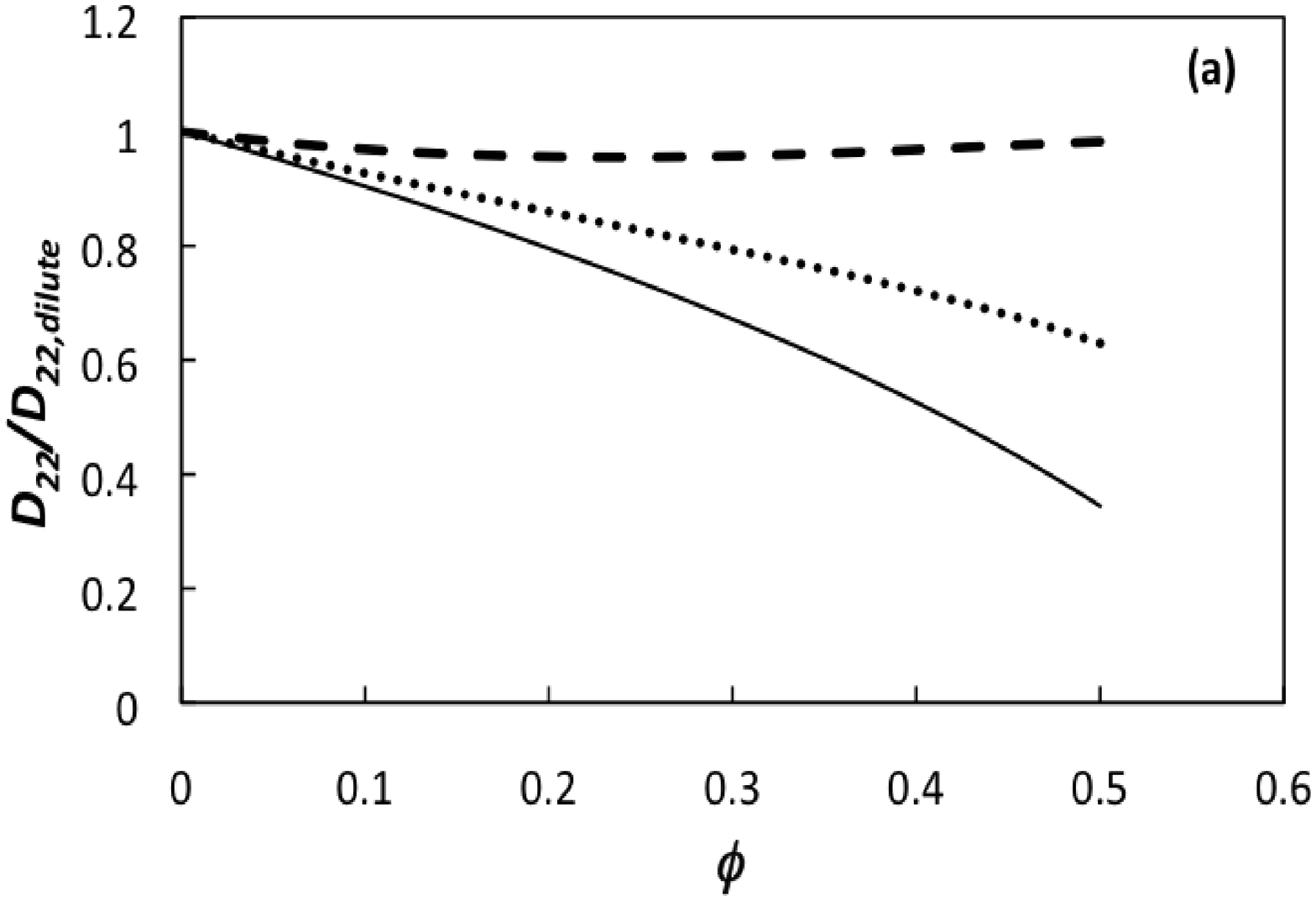,height=2.05in}} &
{\psfig{figure=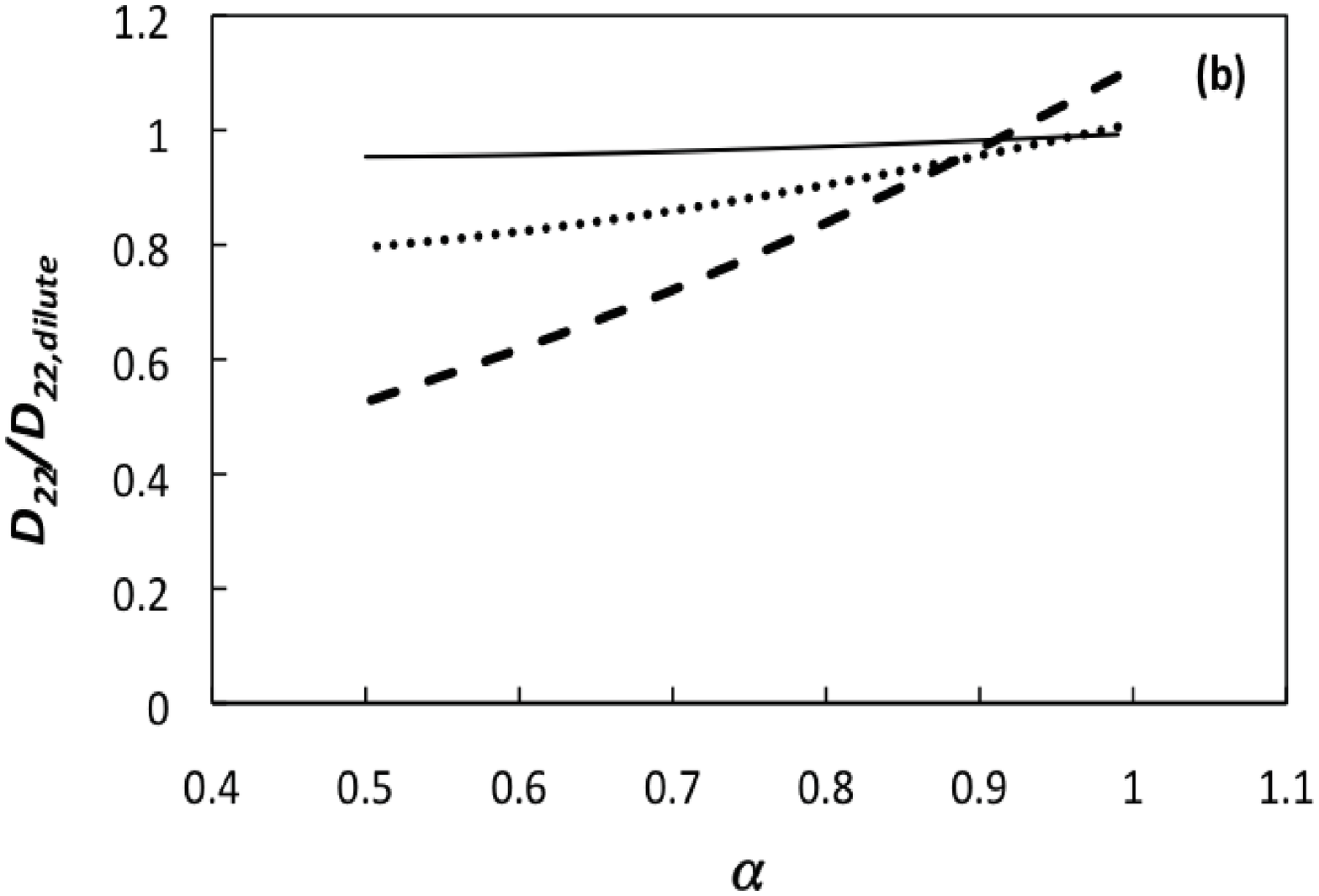,height=2.05in}}
\end{tabular}
\caption{
Mutual diffusion ($D_{22}$):  ratio of moderately dense (GHD) to
dilute (GD) predictions as a function of (a)~overall volume fraction
and (b)~coefficient of restitution. See legends presented in Figure 1.}
\end{figure}
\subsection{Mass flux: Mutual diffusion, thermal diffusion}

The mutual and thermal diffusion coefficients ($D_{ij}$, $D_i^T$) are elements
of the constitutive equation for the mass flux. Based on the identities
given by Eq.\ (\ref{2.5}), the dimensionless mutual diffusion can be
described by two quantities ($D_{11}/D_{11,{\rm dilute}}$ and
$D_{22}/D_{22,{\rm dilute}}$), whereas the dimensionless thermal diffusion can
be described by a single quantity ($D_1^T/D_{1,{\rm dilute}}^T$).
Note that the dilute (GD) mutual and thermal diffusion coefficients
presented by Garz\'{o} and Dufty \cite{GD02} are defined using different
spatial gradients than those used in the dense (GHD)
theory and shown in Eq.\ (\ref{2.1}).  Nonetheless, a conversion
is made such that both dense and dilute theories use the same representations
for the fluxes, namely those shown in Sec.\ \ref{sec2} of this paper,
thereby ensuring an apples-to-apples comparison.

An examination of the dense-to-dilute ratio of the mutual diffusion
coefficient elements (Figures 6 and 7) reveals a more complicated
behavior of these quantities. Because $D_{11,{\rm dilute}}$ approaches zero
for inelastic systems (at $\alpha\sim0.52$), the coefficient of restitution was varied
between 0.6 and 0.99 (Fig.\ 6b). As shown in Figure 6a,
$D_{11}/D_{11,{\rm dilute}}$
is non-monotonic with respect to the volume fraction in less elastic
systems ($\alpha=0.5$), reaching a maximum ratio between the dense and
dilute predictions of 20 at a volume fraction of 0.37. As the system
becomes more elastic, $D_{11}/D_{11,{\rm dilute}}$ shifts from positive to
negative. A change in the sign, as well as magnitude,
between dense and dilute predictions of the mutual diffusion
coefficient $D_{11}$ may provide insight into counter intuitive
species segregation \cite{BRM05,G06,G08,G09}.

The results for $D_{22}/D_{22,{\rm dilute}}$, which are displayed in Figure 7, reveal increasing discrepancies between predictions as volume fraction increases and restitution coefficient decreases. In other words, GHD and GD theories display a larger discrepancy in denser, less elastic systems. For a relatively inelastic and dense system ($\alpha=0.5$ and $\phi=0.5$), the GHD prediction is about half of its dilute counterpart (Fig.\ 7a). However, it is significant
to note the minor differences that exist between the dense and dilute
predictions for the mutual diffusion coefficient $D_{22}$ near the elastic
limit ($\alpha=0.9$) over a range of volume fractions from $\phi=10^{-8}$ to 0.5 (Fig.\ 7a,
$D_{22}/D_{22,{\rm dilute}}\sim1$). GHD and GD theories display a larger discrepancy in denser, less elastic systems. For a relatively inelastic and dense system ($\alpha=0.5$, $\phi=0.4$), the moderately
dense-phase theory prediction is about half of its dilute counterpart
(Fig.\ 7a). Comparing dense- and dilute-phase predictions for the
individual elements $D_{11}$ and $D_{22}$ shows the relative importance
of each contribution to the mutual diffusion. At a moderately low volume
fraction and high coefficient of restitution ($\phi=0.1$, $\alpha=0.9$), the
discrepancies for GHD and GD theory predictions are about 70\% and 5\%
for $D_{11}$ and $D_{22}$, respectively. The dilute theory does not consider the finite size of the particles, which is the main difference between dense and dilute predictions. The discrepancy between $D_{11}$ and $D_{11,\text{dilute}}$ is larger than the discrepancy between $D_{22}$ and $D_{22,\text{dilute}}$ because $D_{11}$ is directly related to the size of species $1$, whereas $D_{22}$ is proportional to the size of species $2$ (recall $\sigma_1/\sigma_2=2$ for the case examined). Neither dilute quantity contains species size, therefore, the self-diffusion coefficient of a relatively large particle compared to its dilute counterpart will be greater than that of its smaller counterpart.

\begin{figure}[htb]
\begin{tabular}{ll}
{\psfig{figure=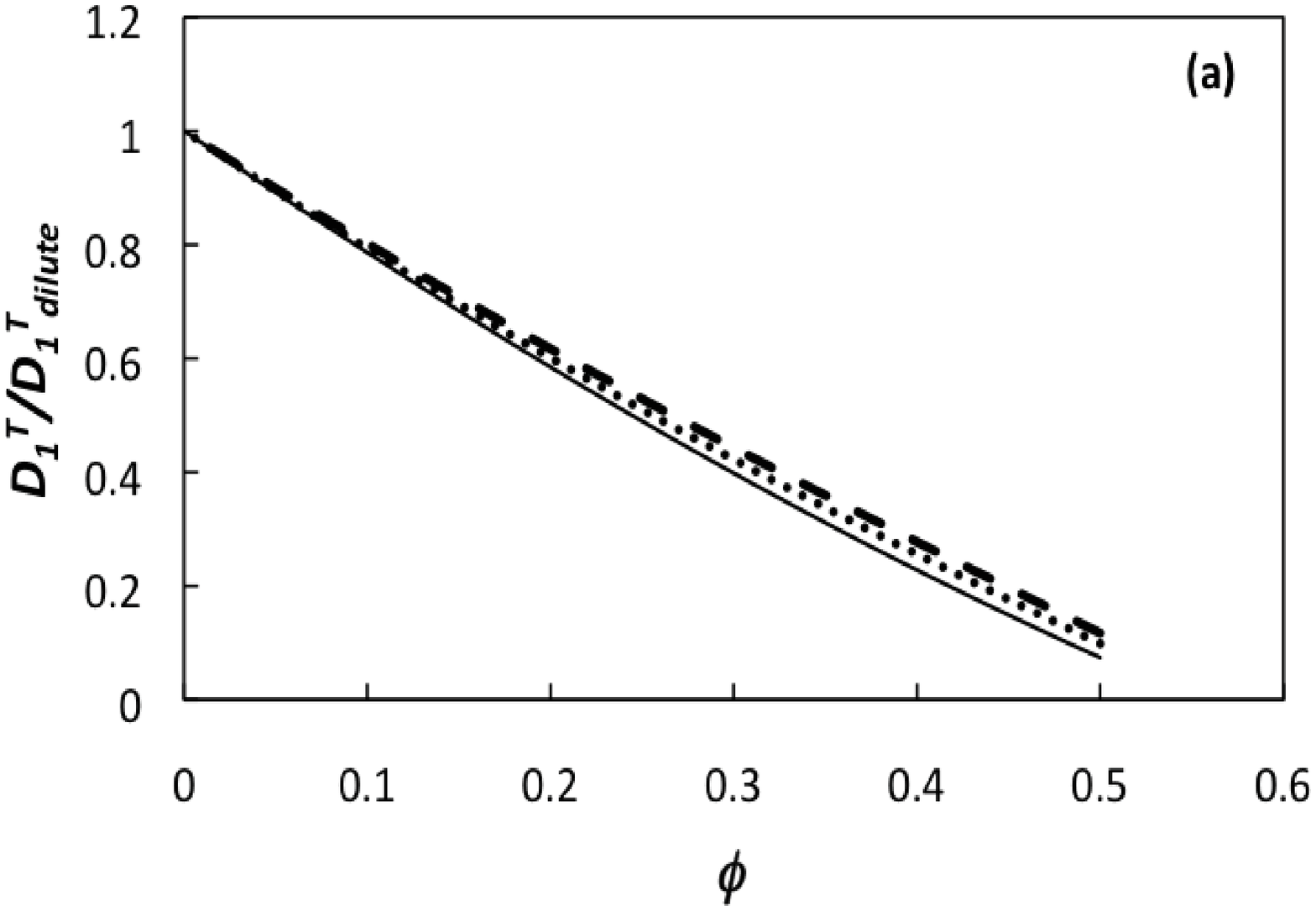,height=2.05in}} &
{\psfig{figure=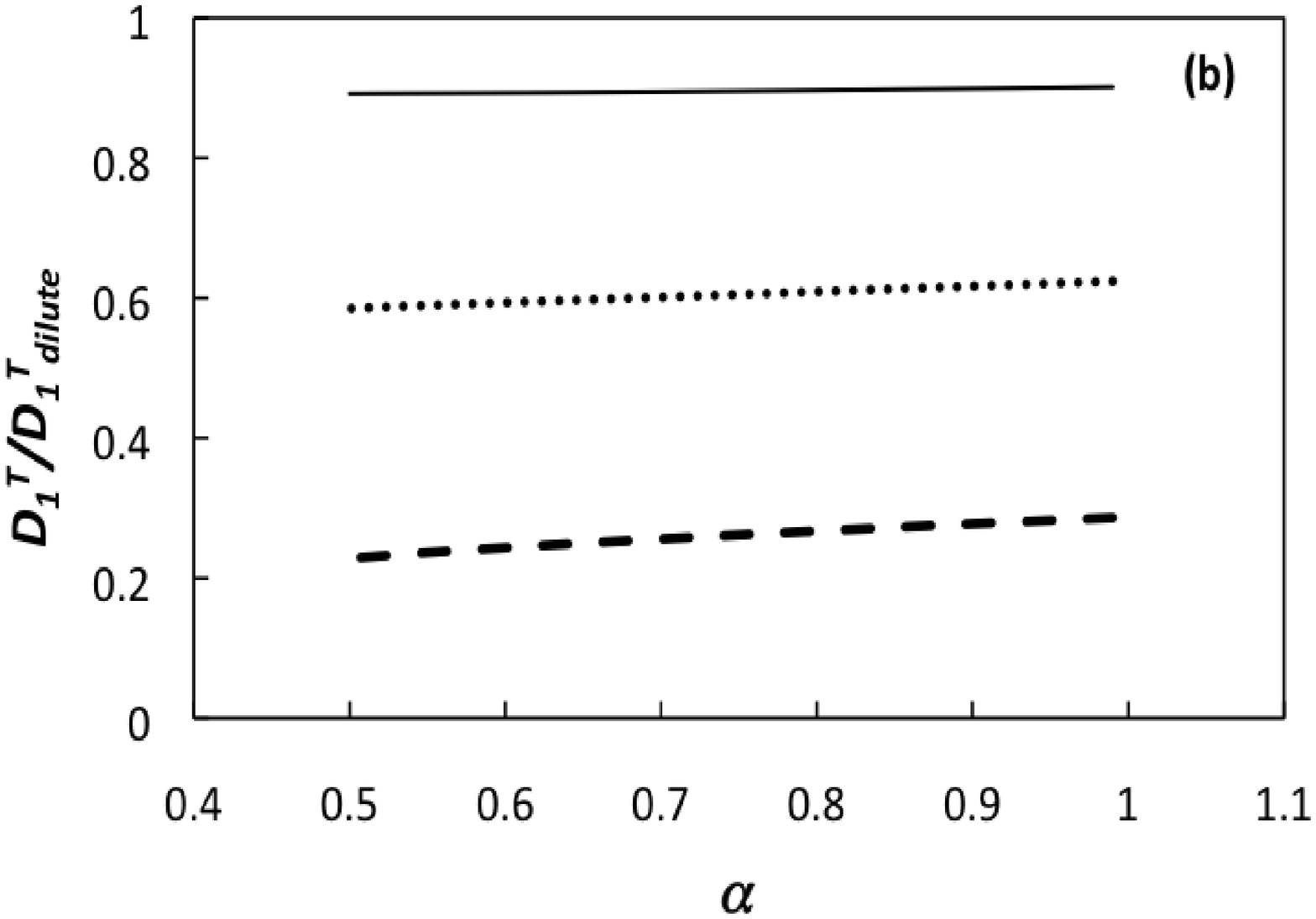,height=2.05in}}
\end{tabular}
\caption{
Thermal diffusion:  ratio of moderately dense (GHD) to dilute
(GD) predictions as a function of (a)~overall volume fraction and
(b)~coefficient of restitution. See legends presented in Figure 1.}
\end{figure}
\begin{figure}[htb]
\begin{tabular}{ll}
{\psfig{figure=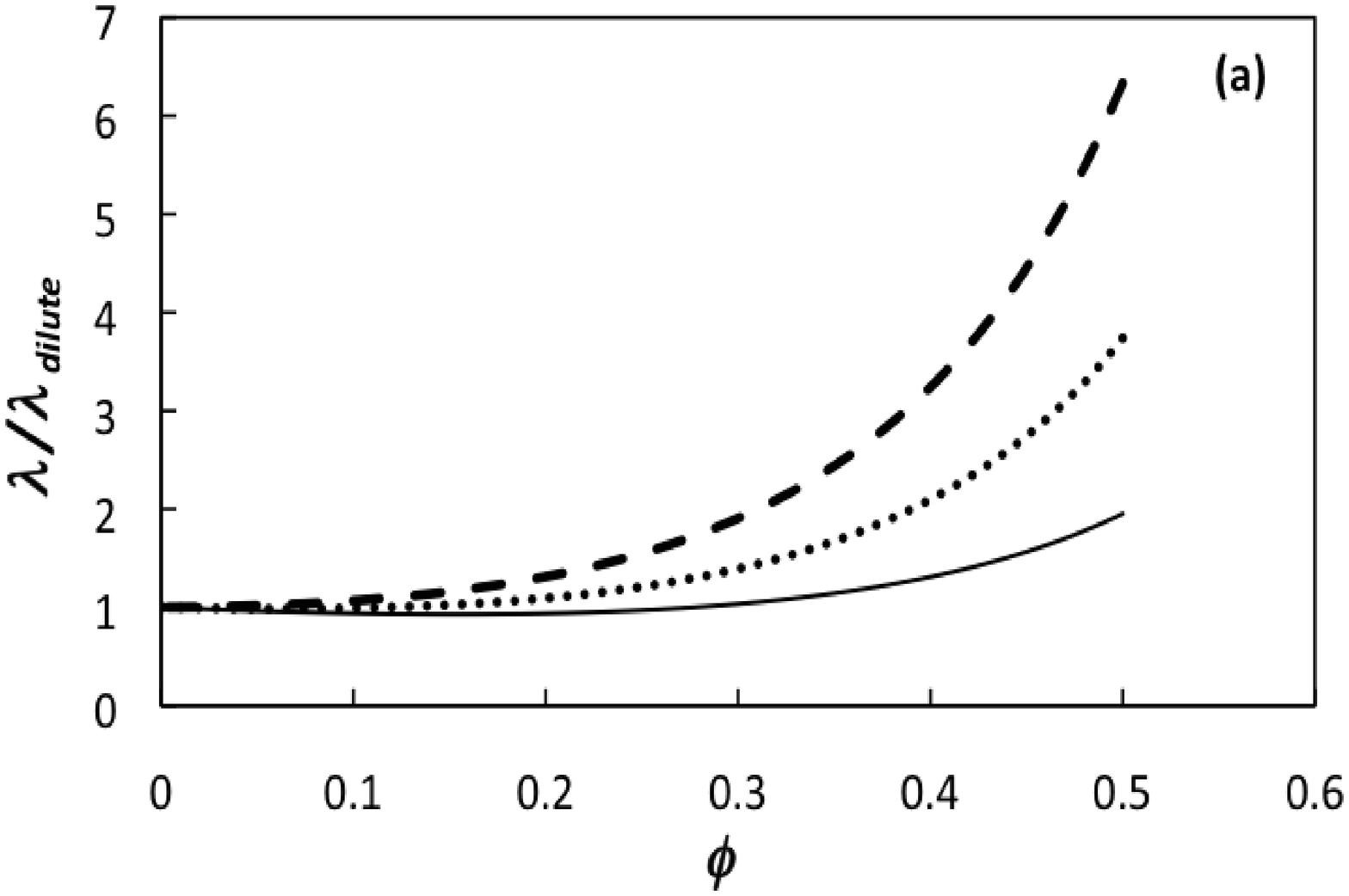,height=2.05in}} &
{\psfig{figure=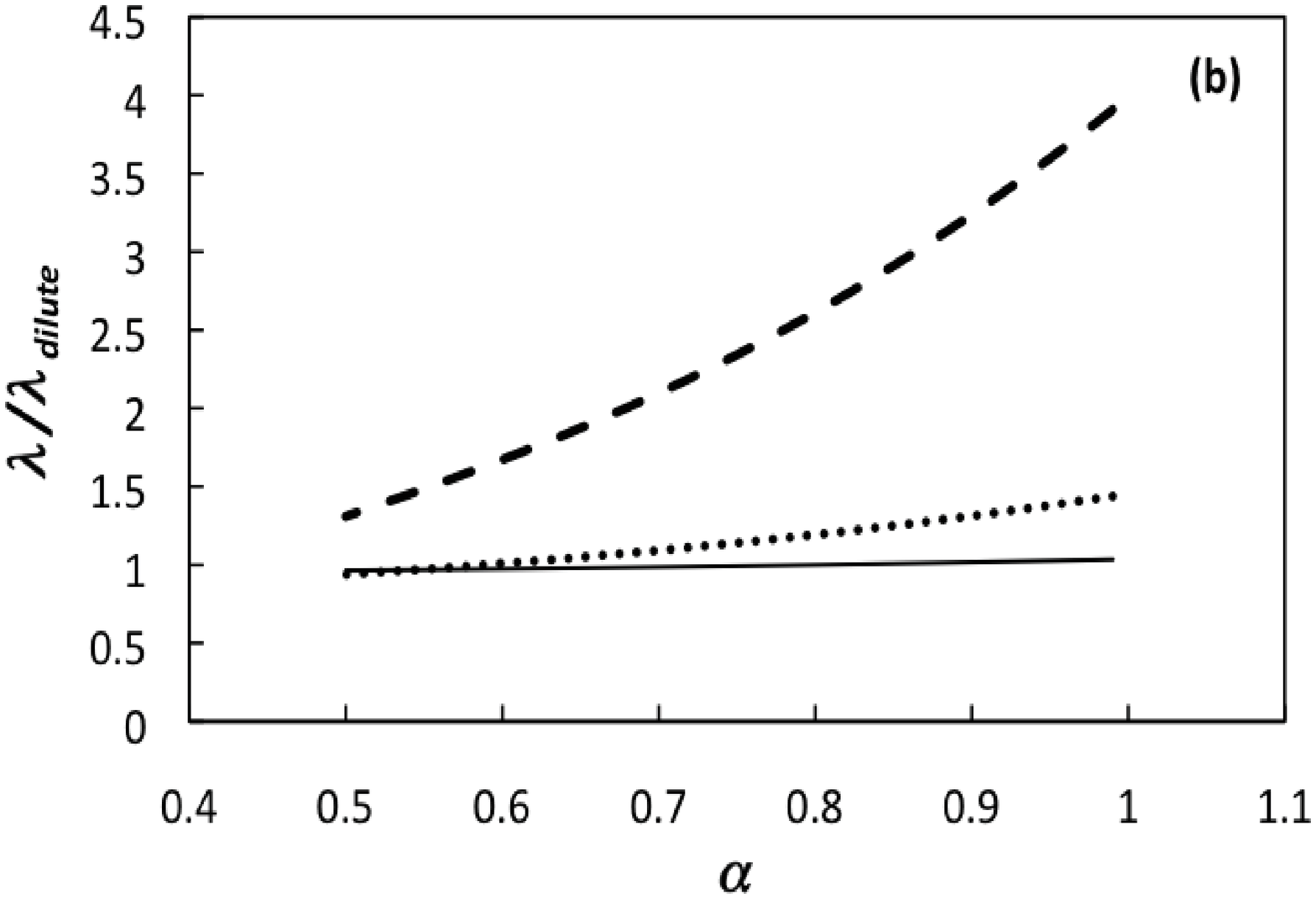,height=2.05in}}
\end{tabular}
\caption{
Thermal Conductivity:  ratio of moderately dense (GHD) to dilute (GD)
predictions as a function of (a)~overall volume fraction and
(b)~coefficient of restitution. See legends presented in Figure 1.}
\end{figure}

The results for thermal diffusion (Figure 8) indicate that the ratio of
dense-to-dilute predictions is extremely sensitive to changes in
volume fraction compared to the coefficient of restitution. These general
trends were also observed in the cooling rate and momentum flux
relations (Figures 1, 3-5). The quantity $D_1^T/D_{1,{\rm dilute}}^T$ is nearly
linear when plotted as a function of volume fraction, regardless the
restitution coefficient (Fig.\ 8a). In a moderately dense system
($\phi=0.4$), the results of Fig.\ 8a indicate that the dilute (GD) theory
prediction of $D_1^T$ is 5 times larger than predicted by GHD theory. At
a much lower volume fraction of 0.1, the dilute (GD) theory
prediction is larger than the moderately dense-phase (GHD) theory
prediction by 20\% (Fig.\ 8a).
\begin{figure}[htp]
\begin{tabular}{ll}
{\psfig{figure=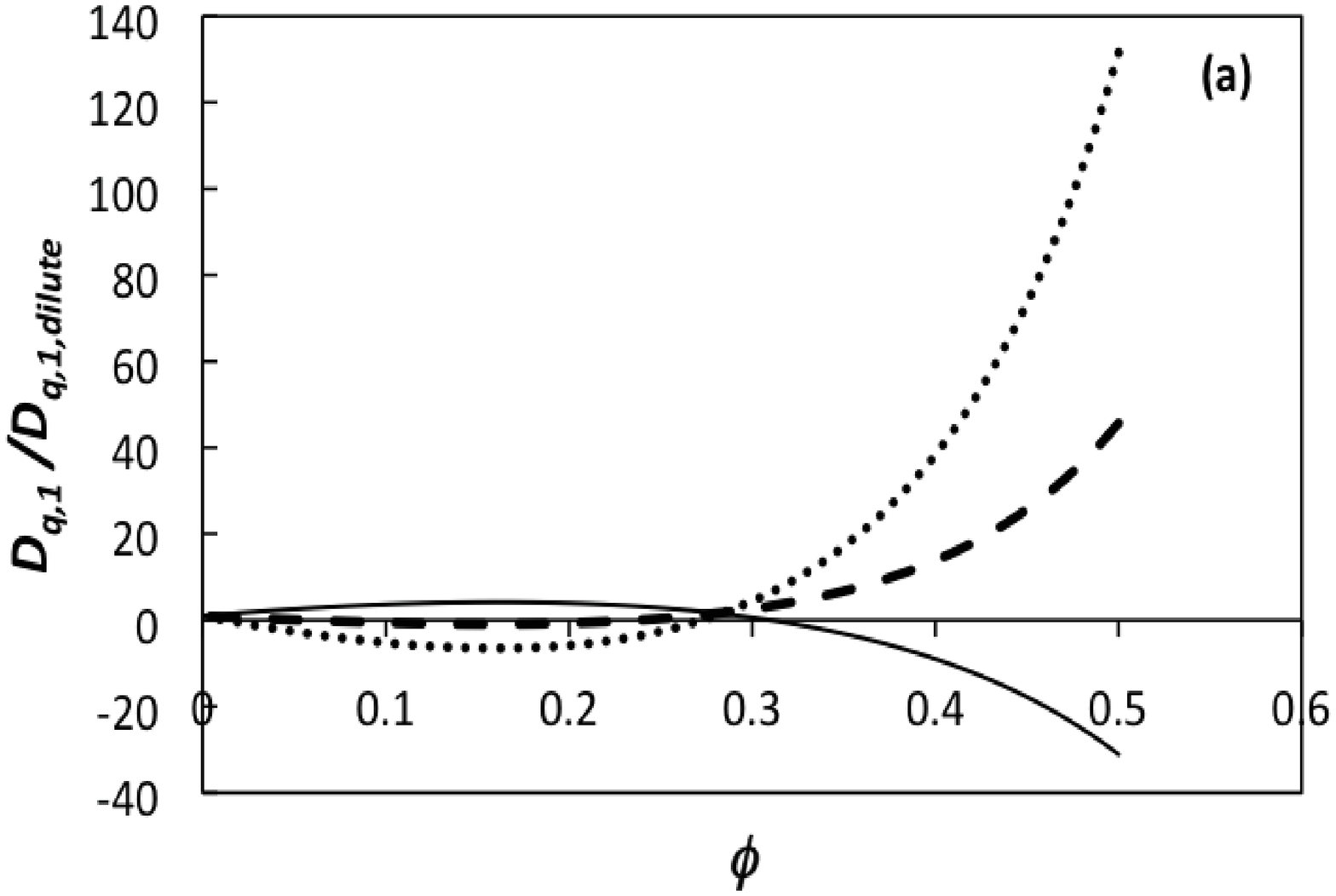,height=2.1in}} &
{\psfig{figure=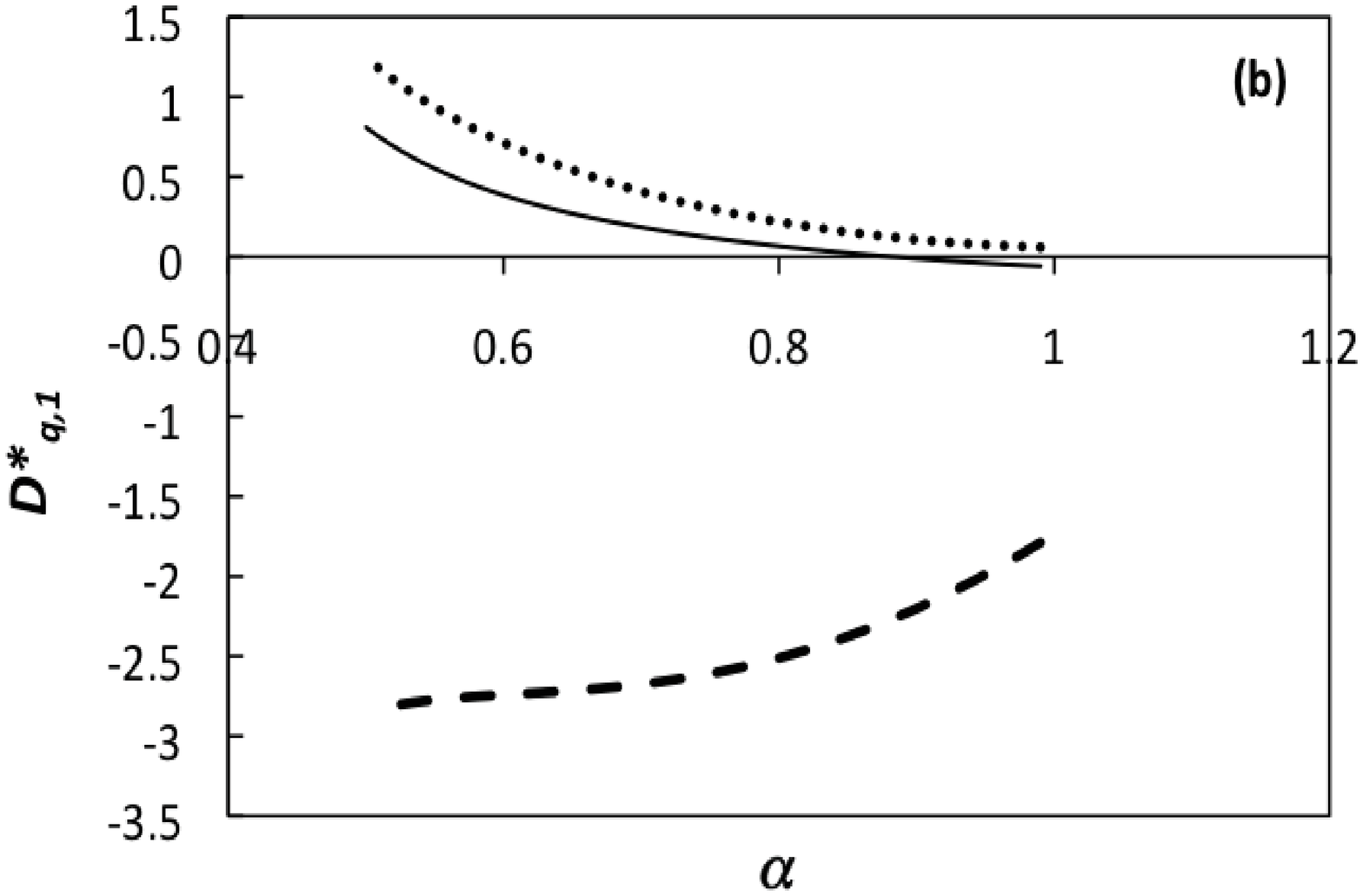,height=2.1in}}\end{tabular}
\centerline{\psfig{figure=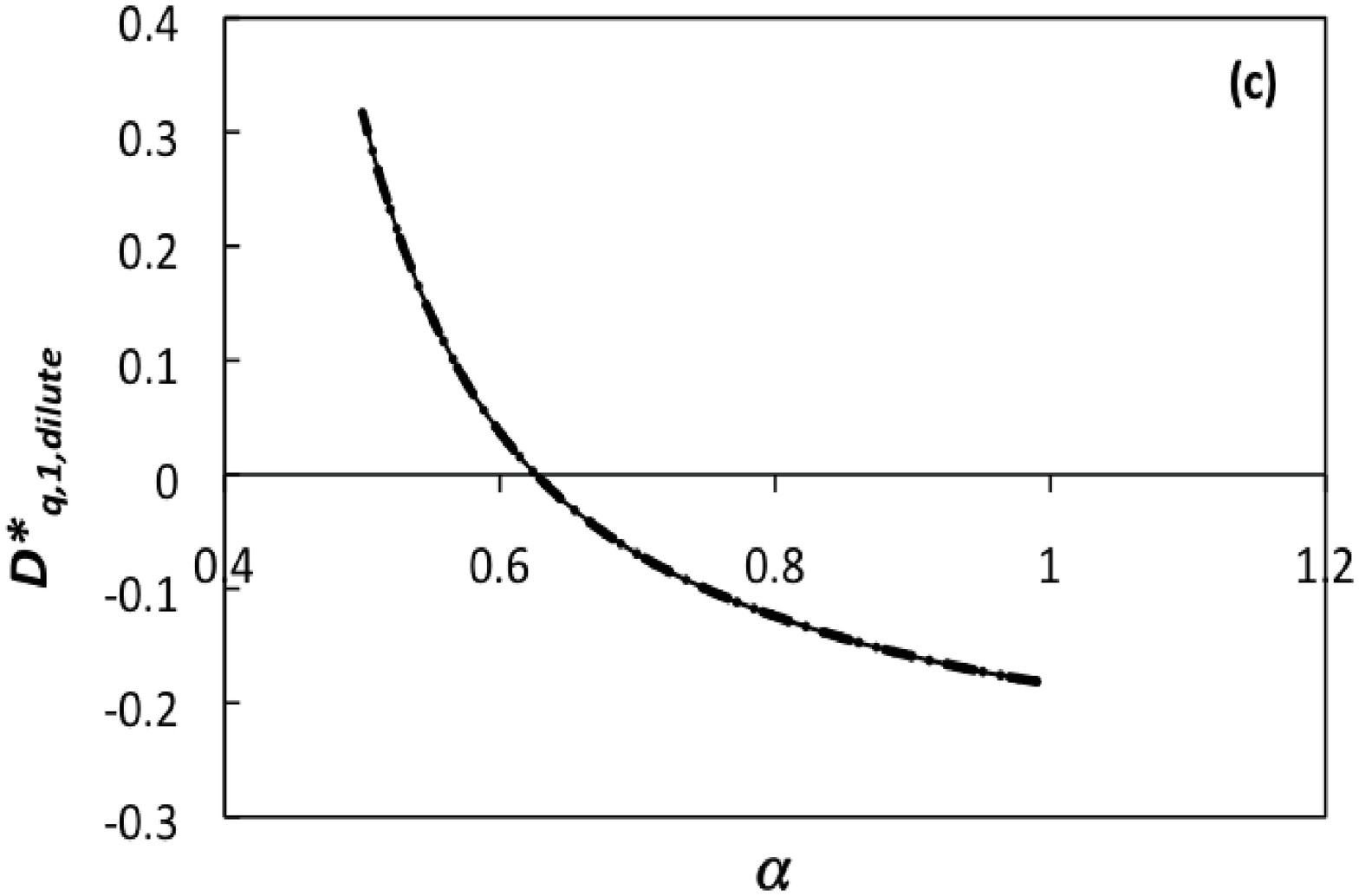,height=2.1in}}
\caption{
Fig.\ 10. Dufour coefficient ($D_{q,1}$):  (a) ratio of moderately dense (
GHD) to dilute (GD) predictions as a function of overall volume fraction
(b) dimensionless moderately dense (GHD) predictions as a function of
coefficient of restitution (c)~dimensionless dilute (GD) predictions as
a function of coefficient of restitution. See legends presented in Figure 1.
The dimensionless inputs are as follows: $m_1/m_2 = 8$, $\sigma_1/\sigma_2 = 2$, and $\phi_1/\phi = 0.5$.}
\end{figure}
\begin{figure}[htb]
\begin{tabular}{ll}
{\psfig{figure=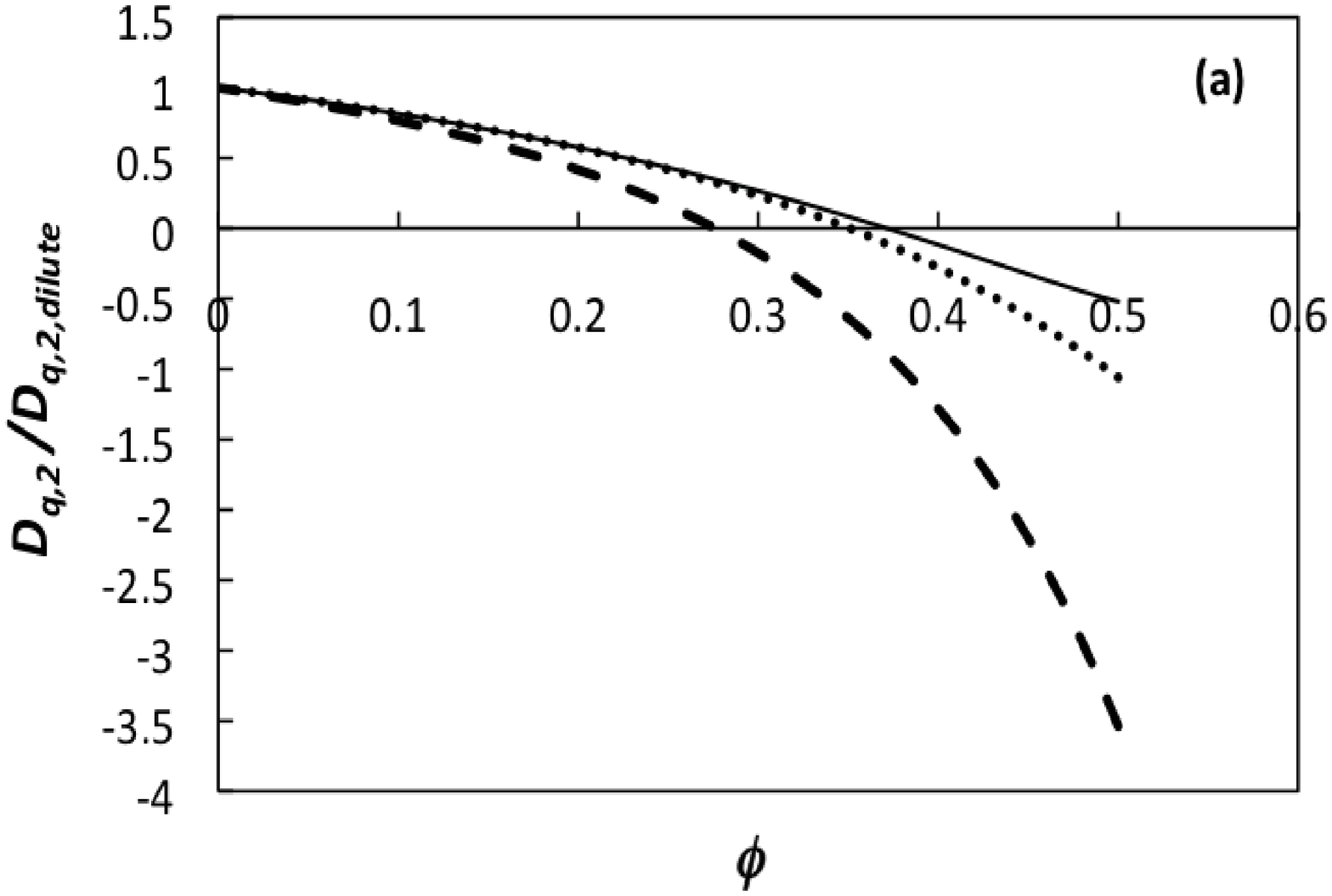,height=2.1in}} &
{\psfig{figure=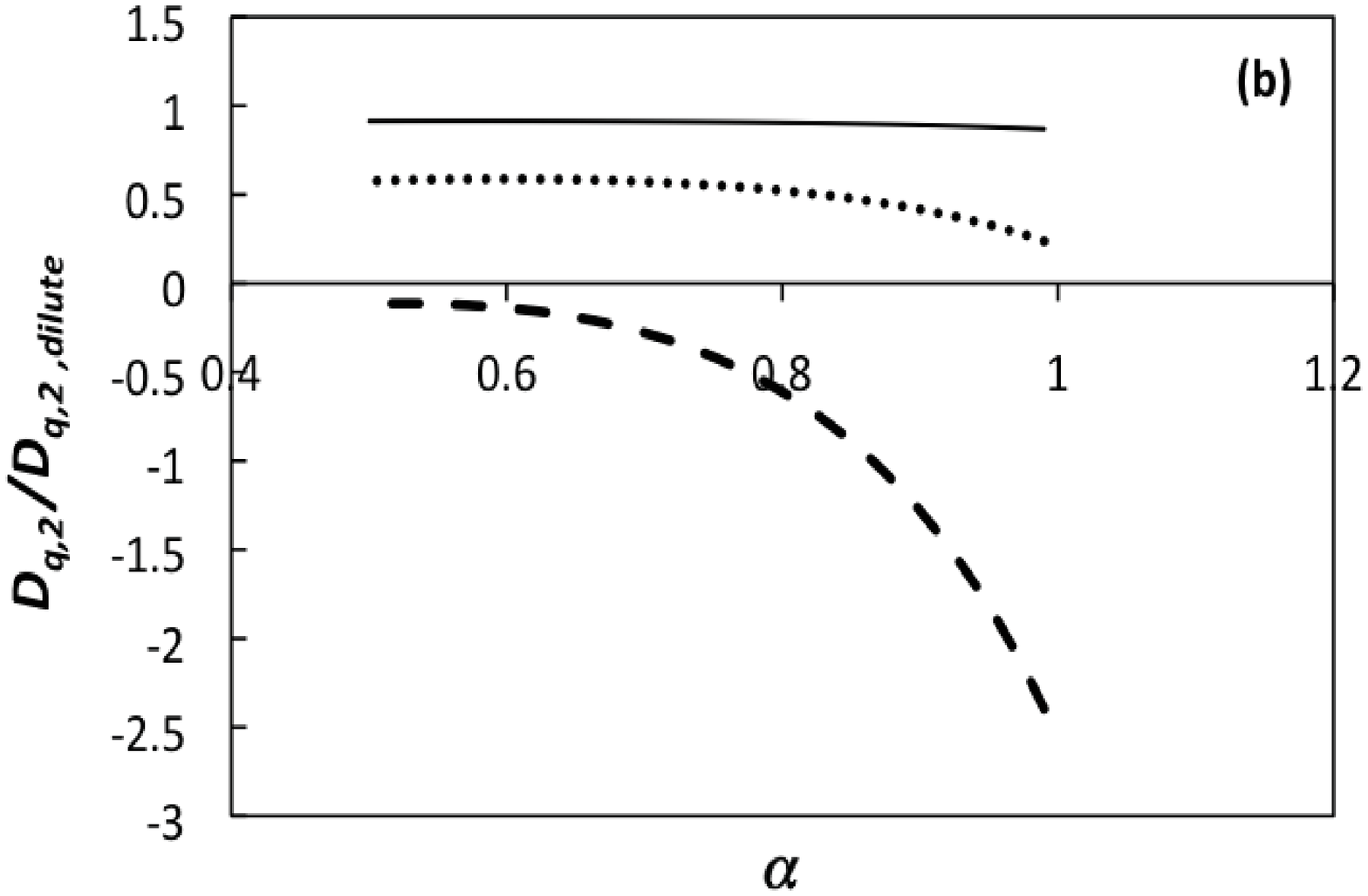,height=2.05in}}
\end{tabular}
\caption{Dufour coefficient ($D_{q,2}$): ratio of moderately dense (GHD) to dilute (GD) predictions as a function of (a) overall volume fraction and (b) coefficient of restitution. See legends presented in Figure 1.}
\end{figure}

\subsection{Heat flux: thermal conductivity, Dufour coefficients}

Heat flux is characterized by the thermal conductivity $\lambda$ and the
Dufour coefficients $D_{q,i}$.
Figure~9 shows that the dense-to-dilute ratio of thermal conductivity increases
monotonically with respect to both volume fraction and
coefficient of restitution. In an elastic, moderately dense system ($\phi\sim 0.4$), results
(Fig.~9a) indicate that the prediction of thermal conductivity from GHD
theory is 4 times larger than that of its dilute (GD) counterpart.
For systems of lower densities ($\phi=0.1$), the discrepancies range from
1\% ($\alpha=0.9$) to 6\% ($\alpha=0.5$) (Fig.\ 9a).

Similar to the mutual diffusion coefficient, the dilute form of the Dufour coefficient takes on a zero value at certain $\alpha$, thereby making the dense-to-dilute value diverge at this value of $\alpha$. Because this value occurs at a practical value of $\alpha=0.63$ (whereas $D_{11,\text{dilute}}$ diverges at $\alpha=0.52$), the dense and dilute predictions of the dimensionless Dufour
coefficient $D_{q,1}^*$ were instead plotted separately against the
coefficient of restitution (Figs.\ 10b and 10c), with the
non-dimensionalization defined in Eq.\ (\ref{2.4}). As expected, the
dilute prediction of the Dufour coefficient is independent of the volume
fraction (Fig. 10c).

The differences in magnitude between dilute and moderately dense
predictions are non-trivial for both $D_{q,1}$ and $D_{q,2}$. More specifically,
the discrepancies that exist between the predictions of $D_{q,1}$ and $D_{q,1,\text{dilute}}$
are up to 2 orders of magnitude in some cases (Figure 10). For a
moderately dense, inelastic system ($\phi=0.5$, $\alpha =0.7$), the
dense-phase prediction is over 100 times greater than its dilute
counterpart (Fig. 10a). Even at a much lower volume
fraction ($\phi=0.01$), the discrepancy between dense and dilute
predictions is at least 30\%. The differences between dense and dilute
predictions of $D_{q,2}$, shown in Figure 11, are less pronounced than
$D_{q,1}$, however, still quite significant. In fact, results indicate
at least a 20\% discrepancy between predictions at a volume fraction
$\phi=0.1$ (Fig.\ 11a). As for the mutual and thermal diffusion coefficients,
the dilute (GD) Dufour coefficient presented in Ref.\ \cite{GD02} is
defined using different spatial gradients than those used in the dense
(GHD) theory and shown in Eq.\ (\ref{2.2}).  As done before, a conversion
has been applied to compare the Dufour coefficients by using the same representation
for the heat flux.

\section{Summary}
\label{sec5}

To date, the understanding of particle segregation within
polydisperse, rapid granular flows is somewhat limited due to a wide
array of complexities that arise during the associated derivation of
continuum theories. As previously mentioned, the two most common
simplifications used in previous theories have been a Maxwellian
velocity distribution and an equipartition of energy. This study
focuses on two particular theories, neither of which assumes the
above conditions. The first was proposed by Garz\'{o} and Dufty \cite{GD02,GMD06}
for binary, dilute mixtures (referred to as GD theory), and the
second was recently proposed by Garz\'{o}, Hrenya and Dufty \cite{GDH07,GHD07} for binary,
moderately dense mixtures (referred to as GHD theory).  In order to
gauge the importance of this dense-phase extension, the transport
coefficients and equations of state predicted by GHD theory were
compared to their dilute counterparts (GD theory). Furthermore, although not the focus of this study, it is worthwhile to mention that the CPU time required to evaluate the dense-phase coefficients was typically three times the requirement for its dilute counterpart.

A systematic comparison was carried out for three different cases
(equal size and different mass, equal mass and different size, and
different size and mass) over a range of mixture parameters (diameter
ratio, mass ratio, and volume fraction ratio), the details of which
are listed in Table~2. Though this study focuses on a case of
different-sized species with the same material density, similar trends were observed for
all other cases analyzed. Results
indicate that transport coefficients and equations of state predicted
by GHD theory are substantially different than those predicted by
dilute (GD) theory. Also, significant
differences between predictions were reported for fairly dilute
systems ($\phi=0.1$). In particular, the discrepancy between predictions
was found to be as large as an order of magnitude. Certain
coefficients, namely the mutual diffusion coefficient $D_{11}$, revealed
that the magnitude and sign were different for the two
theories. Naturally, the level of desired accuracy may vary between users
of the theories.  If, for example, 5\% deviation between the GHD and GD
predictions is deemed acceptable, then the need for a dense-phase
correction is quite evident since the vast majority of
quantities predicted by GHD theory are either larger or smaller than
GD theory predictions by the 5\% limit. Nonetheless, it is worthwhile to point that the
comparison of dense- and dilute-phase predictions for a binary mixture presented here are independent of flow geometry. It is expected that is some flow geometries, one or more of the transport coefficients may dominate, while in other geometries another coefficient(s) may dominate. Such differences are system-dependent and should be taken into account when using the results contained herein.

Given the importance of the dense-phase corrections on the equations
of state and transport coefficients, several follow-on studies are
warranted:  application of the theory to segregating systems in order
to better understand the dominant segregation mechanisms (some previous studies have
been carried in the tracer limit  \cite{G08,G09,GF09}), comparison
with experimental and/or molecular-dynamics simulation data for
purposes of validation, and application of the theory to a continuous
particle size distribution.  It is worthwhile to note that the GHD
theory has been incorporated recently into the open-source, public
MFIX code (https://mfix.netl.doe.gov/) for the case of binary
mixtures, thereby increasing its availability to a wider class of
researchers.

\acknowledgments

J.A.M. and C.M.H. are grateful for the funding support provided by
the Department of Energy (DE-FC26-07NT43098) and the National Science
Foundation (CBET-0318999). The research of V.G. has been supported by the Ministerio de
Educaci\'on y Ciencia (Spain) through grant No. FIS2010-16587, partially financed by
FEDER funds and by the Junta de Extremadura (Spain) through Grant No. GRU10158.

\appendix

\section{Corrections to previous results}
\label{appA}

In this Appendix we explicitly state some changes we have made in the original papers \cite{GDH07} and \cite{GHD07} to correct several errors and/or misprints we have found while working the present manuscript. With these changes, the interested reader can easily obtain the complete set of equations for the mass, heat and momentum fluxes and the cooling rate displayed along Section \ref{sec2}.

Now, we list the changes affecting both papers:

\begin{itemize}

\item In Eq.\ (6.18) of Ref.\ \cite{GDH07}, the term $n_\ell \partial_{n_j}\ln \chi_{i\ell}^{(0)}$ appearing in the second line of the right hand side of this equation must be replaced by $n_j \partial_{n_j}\ln \chi_{i\ell}^{(0)}$. This change affects to Eq.\ (C6) of Ref.\ \cite{GHD07} so that, its left hand side should read
$$
\sum_{\ell=1}^{s}n_\ell \chi_{i\ell}^{(0)}\sigma_{i\ell}^d\left(n_{j}\frac{\partial \ln \chi
_{i\ell }^{(0)}}{\partial n_{j}}+I_{i\ell j}\right)
$$
Equations \eqref{2.7} and \eqref{2.8} can be derived after considering these changes.

\item A factor ``3'' and the diameter $\sigma_{ij}$ are missing in the expression (F25) of Ref.\ \cite{GDH07} in the collisional contribution to the heat flux. Thus, the third and fourth lines of the right hand side of this equation become
$$
\left.
+\frac{24B_{2}}{2+d}n_{i}\left( \frac{2\mu _{ji}}{m_{j}}{\bf
q}_{j}^{k}-(d+2)\frac{T_{i}^{(0)}}{m_{i}m_j}\left( 2\mu _{ij}-\mu _{ji}\right)  {\bf j}_{0j}^{(1)}\right)
+\sigma_{ij}C_{ij}^{T}\nabla \ln T+\sigma_{ij}\sum_{p=1}^{s}C_{ijp}^{T}\nabla \ln n_{p}\right],
$$
where $\mu_{ij}\equiv m_i/(m_i+m_j)$. These changes also affect to Eqs.\ (7.14)-(7.16) of Ref.\ \cite{GDH07} and to Eqs.\ (3.37)-(3.39) of Ref.\ \cite{GHD07}(collisional contributions to the heat flux transport coefficients). Taking into these changes, one gets Eqs.\ \eqref{2.17} and \eqref{2.18} of the present paper.

\item The first line of the right hand side of Eq.\ (3.61) of Ref.\ \cite{GHD07} must be corrected. It should be given by
$$
\overline{d}_{q,ij} =-\frac{d+2}{2}\frac{n_{i}n_{j}T_{i}^{3}}{m_{i}T^{2}} \left( \frac{m_{j}}{\rho
T_{i}}\sum_{\ell =1}^{s}m_\ell\frac{\omega _{i\ell }-\zeta ^{(0)}\delta _{i\ell }}{n_{\ell }T_{\ell }}D_{\ell
j}-\frac{2}{d+2}\frac{m_iT}{n_iT_i^3}\frac{\partial \zeta ^{(0)}}{\partial n_{j}}\lambda _{i}-\frac{1}{dT_{i}}\frac{\partial \ln
T_{i}}{\partial n_{j}}\right)+\cdots
$$
Equation \eqref{a11} of the present paper can be obtained after this change.

\item In Eq.\ (2.16) of Ref.\ \cite{GHD07}, the right hand side should read
$$
\frac{2}{d(d+2)n_i}\overline{e}_{i,D}
$$

\item A minus sign lacks on the second line of Eq.\ (2.19) of Ref.\ \cite{GHD07}. Thus, this line should read
$$
=-\frac{\pi ^{d/2}}{4d\Gamma \left( \frac{d}{2}\right) } \sum_{j=1}^{s}n_{i}n_{j}\chi _{ij}^{(0)}
\cdots
$$

\item In Eq.\ (2.20) of Ref.\ \cite{GHD07}, the term $e_{i,D}$ should be inside the summation sign. Moreover, the partial densities $n_in_j$ must be also included inside the summation sign. Thus, the first line of Eq.\ (2.20) should read
$$
\zeta ^{(1,1)}=\frac{3\pi ^{(d-1)/2}}{4d\Gamma \left( \frac{d}{2} \right)
}\frac{v_{0}^{3}}{nT}\sum_{i=1}^{s}\sum_{j=1}^{s}n_in_j e_{i,D}\sigma _{ij}^{d-1}\chi _{ij}^{(0)}
\cdots
$$
Equation \eqref{5.3} of the present paper can be easily obtained after taking account these changes for the cooling rate.

\item The summation $\sum_{j=1}^s$ is missing on the right hand side of Eq.\ (3.34) of Ref.\ \cite{GHD07}.

\item On the right hand side of Eqs.\ (A28) and (A29) of Ref.\ \cite{GHD07}, the term $1+\alpha_{ij}$
should be changed to $1+\alpha_{ii}$.

\item The ratio $m_\ell/m_i$ on the left hand side of Eq.\ (3.21) of Ref.\ \cite{GHD07} should be removed.

\item In Eq.\ (A12) of Ref.\ \cite{GHD07}, the factor $d+5$ near the end of the second line should be replaced by the factor $d+3$.

\item The right hand side of Eq.\ (3.54) of Ref.\ \cite{GHD07} should read
$$
\lambda^k=\sum_{i=1}^s\; \lambda_i^k=\sum_{i=1}^s\; \lambda_i+\ldots
$$

\end{itemize}

\section{Some explicit expressions}
\label{appB}

The kinetic part of the transport coefficients $D_{q,i}$ and $\lambda$ are given by Eqs.\ (\ref{2.15}) and (\ref{2.16}), respectively. The (dimensionless) Sonine coefficients $\lambda_i^*$ are defined by the matrix equation
\begin{equation}
\label{a3} \left(
\begin{array} {cc}
\gamma_{11}^*-2\zeta^*&\gamma_{12}^*\\
\gamma_{21}^*&\gamma_{22}^*-2\zeta^*
\end{array}
\right)
\cdot
\left(
\begin{array}{c}
\lambda_1^*\\
\lambda_2^*
\end{array}
\right)
=\left(
\begin{array}{c}
\overline{\lambda}_1^*\\
\overline{\lambda}_2^*
\end{array}
\right),
\end{equation}
where
\begin{equation}
\label{a4}
\overline{\lambda}_i^*=\frac{m_1+m_2}{m_i}x_i\gamma_i^2\sum_{j=1}^2\left(\delta_{ij}-\frac{\omega_{ij}^*-
\zeta^*\delta_{ij}}{x_j\gamma_j}D_j^{*T}+\frac{\pi^{d/2}}{d(d+2)\Gamma \left(\frac{d}{2}\right)}
n\sigma_2^d M_{ij}x_j(\sigma_{ij}/\sigma_2)^d\chi_{ij}\frac{\gamma_j}{\gamma_i}A_{ij}\right).
\end{equation}
The expressions of the (reduced) collision frequencies
$\gamma_{ij}^*$ and $\omega_{ij}^*$
can be found in the Appendix A of Ref.\ \cite{GHD07}. Moreover, in Eq.\ (A2) we have introduced the quantity
\begin{equation}
A_{ij}=(d+2)(M_{ij}^{2}-1)+(2d-5-9\alpha_{ij})M_{ij}M_{ji}+(d-1+3\alpha _{ij}+6\alpha _{ij}^{2})M
_{ji}^{2}+6\frac{m_{i}T_j}{m_{j}T_i}M_{ji}^{2}(1+\alpha _{ij})^{2}.  \label{a5}
\end{equation}
The solution to Eq.\ ({A1}) is elementary and gives
\begin{equation}
\label{a6}
\lambda_1^*=\frac{(\gamma_{22}^*-2\zeta^*)\overline{\lambda}_1^*-\gamma_{12}^*\overline{\lambda}_2^*}
{4\zeta^{*2}-2(\gamma_{11}^*+\gamma_{22}^*)\zeta^*-\gamma_{12}^*\gamma_{21}^*+\gamma_{11}^*\gamma_{22}^*}, \quad
\lambda_2^*=\frac{(\gamma_{11}^*-2\zeta^*)\overline{\lambda}_2^*-\gamma_{21}^*\overline{\lambda}_1^*}
{4\zeta^{*2}-2(\gamma_{11}^*+\gamma_{22}^*)\zeta^*-\gamma_{12}^*\gamma_{21}^*+\gamma_{11}^*\gamma_{22}^*}.
\end{equation}
With these results the kinetic part $\lambda^{k*}$ can be written as
\begin{equation}
\label{a6.1}
\lambda^{k*}=\frac{\overline{\lambda}_1^*(\gamma_{22}^*-2\zeta^*-\gamma_{21}^*)+
\overline{\lambda}_2^*(\gamma_{11}^*-2\zeta^*-\gamma_{12}^*)}
{4\zeta^{*2}-2(\gamma_{11}^*+\gamma_{22}^*)\zeta^*-\gamma_{12}^*\gamma_{21}^*+\gamma_{11}^*\gamma_{22}^*}
+\left(\frac{\gamma_1}{M_{12}}-\frac{\gamma_2}{M_{21}}\right)D_1^{T*}.
\end{equation}

The kinetic part of the transport coefficients $D_{q,i}^{k*}$ is given
in terms of the Sonine coefficients $d_{q,ij}^*$. By using matrix notation, the coupled set of four equations for the coefficients
\begin{equation}
\label{a7}
\left\{d_{q,11}^*, d_{q,12}^*, d_{q,21}^*, d_{q,22}^*\right\}
\end{equation}
can be written as
\begin{equation}
\label{a8}
\Lambda_{\mu \mu'}X_{\mu'}=Y_\mu.
\end{equation}
Here, $X_\mu$ is the column matrix defined by the set (A6) and $\Lambda_{\mu \mu'}$ is the square matrix
\begin{equation}
\label{a9}
\Lambda_{\mu\mu'}=
\left(
\begin{array}{cccc}
\gamma_{11}^*-\frac{3}{2}\zeta^*&0&\gamma_{12}^*&0\\
0&\gamma_{11}^*-\frac{3}{2}\zeta^*&0&\gamma_{12}^*\\
\gamma_{21}^*&0&\gamma_{22}^*-\frac{3}{2}\zeta^*&0\\
0&\gamma_{21}^*&0&\gamma_{22}^*-\frac{3}{2}\zeta^*
\end{array}
\right).
\end{equation}
The column matrix ${\sf Y}$ is
\begin{equation}
\label{a10} {\bf Y}=\left(
\begin{array}{c}
\overline{d}_{q,11}^*\\
\overline{d}_{q,12}^*\\
\overline{d}_{q,21}^*\\
\overline{d}_{q,22}^*\\
\end{array}
\right),
\end{equation}
where
\begin{eqnarray}
\label{a11}
\overline{d}_{q,ij}^*&=&\frac{m_1+m_2}{m_i}x_i\gamma_in_j\frac{\partial \gamma_i}{\partial n_j}-
\frac{m_1+m_2}{m_i}x_ix_j\gamma_i^2\sum_{\ell=1}^2\frac{\omega_{i\ell}^*-
\zeta^*\delta_{i\ell}}{x_\ell\gamma_\ell}D_{\ell j}^{*}
+\frac{n_j}{\nu_0}\frac{\partial \zeta^{(0)}}{\partial n_j}\lambda_i^*
\nonumber\\
& & +\frac{\pi^{d/2}}{d(d+2)\Gamma \left(\frac{d}{2}\right)}
\frac{m_1+m_2}{m_i}x_i\gamma_i^2n\sigma_2^d\sum_{\ell=1}^2
M_{\ell i}x_\ell(\sigma_{i\ell}/\sigma_2)^d\chi_{i\ell}(1+\alpha_{i\ell})
\nonumber\\
& & \times
\left\{\left[\delta_{j\ell}+\frac{1}{2}
\left(n_j\frac{\partial \chi_{i\ell}}{\partial n_j}+I_{i\ell j}\right)\right]B_{i\ell}+
\frac{m_i}{m_\ell \gamma_i}n_j\frac{\partial \gamma_\ell}{\partial n_j}A_{i\ell}\right\}.
\end{eqnarray}
In Eq.\ (A10), $A_{ij}$ is defined by Eq.\ (\ref{a5}) and $B_{ij}$ is given by
\begin{eqnarray}
B_{ij}&=&(d+8)M _{ij}^{2}+(7+2d-9\alpha_{ij})M_{ij}M_{ji}+(2+d+3\alpha _{ij}^{2}-3\alpha
_{ij})M_{ji}^{2}+3M_{ji}^{2}(1+\alpha_{ij})^{2}\frac{m_i^2T_{j}^{2}}{m_{j}^{2}T_i^2}\nonumber\\
& & +\left[
(d+2)M_{ij}^{2}+(2d-5-9\alpha_{ij})M_{ij}M_{ji}+(d-1+3\alpha
_{ij}+6\alpha _{ij}^{2})M_{ji}^{2}\right] \frac{m_iT_{j}}{m_{j}T_i}
\nonumber\\
& & -(d+2)\left(1+ \frac{m_iT_{j}}{m_{j}T_i}\right).\label{a12}
\end{eqnarray}
The solution to Eq.\ (A7) provides the expressions of $d_{q,ij}^*$. The result is
\begin{equation}
\label{a13}
d_{q,11}^*=\frac{4\overline{d}_{q,21}^*\gamma_{12}^*-4\overline{d}_{q,11}^*\gamma_{22}^*+6
\overline{d}_{q,11}^*\zeta^*}{4\gamma_{12}^*\gamma_{21}^*+(2\gamma_{11}^*-3\zeta^*)(3\zeta^*-2\gamma_{22}^*)},
\quad
d_{q,12}^*=\frac{4\overline{d}_{q,22}^*\gamma_{12}^*-4\overline{d}_{q,12}^*\gamma_{22}^*+6
\overline{d}_{q,12}^*\zeta^*}{4\gamma_{12}^*\gamma_{21}^*+(2\gamma_{11}^*-3\zeta^*)(3\zeta^*-2\gamma_{22}^*)}.
\end{equation}
The kinetic part $D_{q,1}^{k*}$ can be easily obtained when one takes into account Eqs.\ (\ref{2.15}) and (A10). The result is
\begin{equation}
\label{a14}
D_{q,1}^{k*}=\frac{4\gamma_{12}^*(\overline{d}_{q,21}^*+\overline{d}_{q,22}^*)+
2(\overline{d}_{q,11}^*+\overline{d}_{q,12}^*)(3\zeta^*-2\gamma_{22}^*)}
{4\gamma_{12}^*\gamma_{21}^*+(2\gamma_{11}^*-3\zeta^*)(3\zeta^*-2\gamma_{22}^*)}+
\left(\frac{\gamma_1}{M_{12}}-\frac{\gamma_2}{M_{21}}\right)
x_1 D_{11}^*.
\end{equation}
The expressions of $d_{q,22}^*$, $d_{q,21}^*$ and $D_{q,2}^{k*}$ can be obtained from Eqs.\ (\ref{a13}) and (\ref{a14}), respectively, by interchanging 1$\leftrightarrow$2.

In order to get the dependence of the transport coefficients on the parameters of the system,
one needs to know the explicit forms of $\chi_{ij}$ and $\mu_i$. For hard disks ($d=2$), a good approximation
for the pair correlation function $\chi_{ij}$ is \cite{JM87}
\begin{equation}
\label{a15}
\chi_{ij}=\frac{1}{1-\phi}+\frac{9}{16}\frac{\phi}{(1-\phi)^2}\frac{\sigma_i\sigma_jM_1}{\sigma_{ij}M_2},
\end{equation}
where
\begin{equation}
\label{a16}
M_n=\sum_{s=1}^2\; x_s \sigma_s^n.
\end{equation}
The expression of the chemical potential $\mu_i$ of the species $i$
consistent with the approximation (\ref{a15}) is \cite{S08}
\begin{eqnarray}
\label{a17}
\frac{\mu_i}{T}&=&\ln (\lambda_i^2n_i)-\ln (1-\phi)+\frac{M_1}{4M_2}\left[\frac{9\phi}{1-\phi}+\ln (1-\phi)\right]\sigma_i
\nonumber\\
& &
-\frac{1}{8}\left[\frac{M_1^2}{M_2^2}\frac{\phi(1-10\phi)}{(1-\phi)^2}-
\frac{8}{M_2}\frac{\phi}{1-\phi}+\frac{M_1^2}{M_2^2}\ln (1-\phi)\right]\sigma_i^2,
\end{eqnarray}
where $\lambda_i(T)$ is the (constant) de Broglie's thermal wavelength \cite{RG73}. In the case of hard spheres ($d=3$), we take for the pair correlation function
$\chi_{ij}$ the following approximation \cite{B70}
\begin{equation}
\label{a18}
\chi_{ij}=\frac{1}{1-\phi}+\frac{3}{2}\frac{\phi}{(1-\phi)^2}\frac{\sigma_i\sigma_jM_2}{\sigma_{ij}M_3}
+\frac{1}{2}\frac{\phi^2}{(1-\phi)^3}\left(\frac{\sigma_i\sigma_jM_2}{\sigma_{ij}M_3}\right)^2.
\end{equation}
The chemical potential consistent with (\ref{a18}) is \cite{RG73}
\begin{eqnarray}
\label{a19}
\frac{\mu_i}{T}&=&\ln (\lambda_i^3n_i)-\ln (1-\phi)+3\frac{M_2}{M_3}\frac{\phi}{1-\phi}\sigma_i
+3\left[\frac{M_2^2}{M_3^2}\frac{\phi}{(1-\phi)^2}+
\frac{M_1}{M_3}\frac{\phi}{1-\phi}+\frac{M_2^2}{M_3^2}\ln (1-\phi)\right]\sigma_i^2\nonumber\\
& & -\left[\frac{M_2^3}{M_3^3}\frac{\phi(2-5\phi+\phi^2)}{(1-\phi)^3}-3
\frac{M_1M_2}{M_3^2}\frac{\phi^2}{(1-\phi)^2}-\frac{1}{M_3}\frac{\phi}{1-\phi}
+2\frac{M_2^3}{M_3^3}\ln (1-\phi)\right]\sigma_i^3.
\end{eqnarray}

\section{Expressions for a low-density granular binary mixture}
\label{appC}

In this Appendix we include the explicit expressions of the transport coefficients and the cooling rate for a \emph{dilute} binary granular mixture. These expressions can be easily obtained from the results derived in Sec.\ \ref{sec2} for a moderately dense binary mixture by taking the limit $n\sigma_2^d\to 0$.

\subsection{Mass and heat flux transport coefficients}

The expressions of the reduced coefficients $D_{1}^{T*}$, $D_{11}^*$,
and $D_{12}^*$ are given by
\begin{equation}
D_{1}^{T*}=\left(\nu_D^*-\zeta^*\right)^{-1}\left(x_1\gamma_1-\frac{
\rho_1}{\rho}\right),  \label{c1}
\end{equation}
\begin{equation}
\label{c2}
D_{11}^*=\left(\nu_D^*-\frac{1}{2}\zeta^*\right)^{-1}\left(\frac{D_{1}^{T*}}
{x_1\nu_0}n_1\frac{\partial \zeta^{(0)}}{\partial n_1}-\frac{\rho_1}{\rho }
+\gamma_1+n_1\frac{\partial \gamma_1}{\partial n_1}\right),
\end{equation}
\begin{equation}
\label{c3}
D_{12}^*=\left(\nu_D^*-\frac{1}{2}\zeta^*\right)^{-1}\left(\frac{D_{1}^{T*}}
{x_2\nu_0}n_2\frac{\partial \zeta^{(0)}}{\partial n_2}-\frac{\rho_1}
{\rho}+n_1
\frac{\partial \gamma_1}{\partial n_2}\right),
\end{equation}
where $\nu_D^*$ is given by Eq.\ \eqref{2.9} with $\chi_{12}=1$. Upon deriving Eqs.\ \eqref{c1}--\eqref{c3}, use has been made of the identities $p^*=1$ and $\partial p/\partial n_i=T$.

The collisional transfer contributions to the heat flux transport coefficients vanish in the low density limit ($n\sigma_2^d\to 0$). Thus, only their kinetic contributions must be considered. In dimensionless forms, the thermal conductivity  $\lambda^*$ and the Dufour  $D_{q,i}^*$ coefficients are
\begin{equation}
\label{c4}
\lambda^{*}=\lambda_1^*+\lambda_2^*+\left(\frac{\gamma_1}{M_{12}}-\frac{\gamma_2}{M_{21}}\right)D_1^{T*},
\end{equation}
\begin{equation}
\label{c5}
D_{q,1}^{*}=d_{q,11}^*+d_{q,21}^*+\left(\frac{\gamma_1}{M_{12}}-\frac{\gamma_2}{M_{21}}\right)
x_1 D_{11}^*, \quad
D_{q,2}^{*}=d_{q,22}^*+d_{q,12}^*+\left(\frac{\gamma_1}{M_{12}}-\frac{\gamma_2}{M_{21}}\right)
x_2 D_{12}^*,
\end{equation}
where the coefficients $\lambda_i^*$ and $d_{q,ij}^*$ are given by Eqs.\ \eqref{a6} and \eqref{a13}, respectively, with
\begin{equation}
\label{c6}
\overline{\lambda}_i^*=\frac{m_1+m_2}{m_i}x_i\gamma_i^2\sum_{j=1}^2\left(\delta_{ij}-\frac{\omega_{ij}^*-
\zeta^*\delta_{ij}}{x_j\gamma_j}D_j^{*T}\right),
\end{equation}
\begin{equation}
\label{c7}
\overline{d}_{q,ij}^*=\frac{m_1+m_2}{m_i}x_i\gamma_in_j\frac{\partial \gamma_i}{\partial n_j}-
\frac{m_1+m_2}{m_i}x_ix_j\gamma_i^2\sum_{\ell=1}^2\frac{\omega_{i\ell}^*-
\zeta^*\delta_{i\ell}}{x_\ell\gamma_\ell}D_{\ell j}^{*}
+\frac{n_j}{\nu_0}\frac{\partial \zeta^{(0)}}{\partial n_j}\lambda_i^*.
\end{equation}
The dilute forms of the collision frequencies $\omega_{ij}^*$ and $\gamma_{ij}^*$ can be obtained from their dense counterparts (Appendix A of Ref.\ \cite{GHD07}) by simply taking $\chi_{ij}=1$.

\subsection{Pressure tensor}

In the low-density limit, the hydrostatic pressure $p=nT$, the bulk viscosity $\kappa=0$ and the shear viscosity $\eta$ has only kinetic contributions. It is given by $\eta=(p/\nu_0)\eta^*$ where $\eta^*=\eta_1^*+\eta_2^*$. The partial contributions $\eta_i^*$ are given by Eqs.\ \eqref{4.6} and \eqref{4.6.0} with
\begin{equation}
\label{c8}
\overline{\eta}_1=x_1\gamma_1, \quad \overline{\eta}_2=x_2\gamma_2.
\end{equation}
As before, the (reduced) collision frequencies $\tau_{ij}^*$ can be easily obtained from their corresponding dense forms by considering $\chi_{ij}=1$.

\subsection{Cooling rate}

The first-order contribution $\zeta_u$ to the cooling rate $\zeta$ vanishes in the dilute case [see Eqs.\ \eqref{5.1}--\eqref{5.5}]. The dilute expression for the coefficient $\zeta^{(0)}=\zeta_i^{(0)}$ can be obtained from Eq.\ \eqref{4.9} by taking $\chi_{ij}=1$.

\end{document}